\documentclass[aps,prl,preprint,superscriptaddress]{revtex4-1}

\usepackage{amsmath}
\usepackage{fixmath}
\usepackage{accents}
\usepackage{graphicx}% 
\usepackage[capitalise]{cleveref}
\usepackage{amsfonts}

\allowdisplaybreaks

\newcommand{\mb}{\mathbold}

\begin{document}
	
	\title{Second harmonic generation in graphene dressed by a strong terahertz field}
	
	\date{\today}
	
	\author{ Mikhail Tokman}

\affiliation{ Institute of Applied Physics, Russian Academy of Sciences, Nizhny Novgorod, 603950, Russia }

\author{ Sergei B. Bodrov}

\affiliation{ Institute of Applied Physics, Russian Academy of Sciences, Nizhny Novgorod, 603950, Russia }

\author{ Yuri A. Sergeev}

\affiliation{ Institute of Applied Physics, Russian Academy of Sciences, Nizhny Novgorod, 603950, Russia }

\author{ Alexei I. Korytin}

\affiliation{ Institute of Applied Physics, Russian Academy of Sciences, Nizhny Novgorod, 603950, Russia }

\author{ Ivan Oladyshkin}

\affiliation{ Institute of Applied Physics, Russian Academy of Sciences, Nizhny Novgorod, 603950, Russia }

	\author{ Yongrui Wang}

\affiliation{ Department of Physics and Astronomy, Texas A\&M University,
College Station, TX, 77843 USA}

\author{ Alexey Belyanin}

\affiliation{ Department of Physics and Astronomy, Texas A\&M University,
College Station, TX, 77843 USA}

\author{ Andrei N. Stepanov}

\affiliation{ Institute of Applied Physics, Russian Academy of Sciences, Nizhny Novgorod, 603950, Russia }

	\begin{abstract}
	We observe enhanced second-harmonic generation  in monolayer graphene in the presence of an ultra-strong terahertz field pulse with a peak amplitude of 250 kV/cm. 	This is a strongly nonperturbative regime of light-matter interaction in which particles get accelerated to energies exceeding the initial Fermi energy of 0.2 eV over a timescale of a few femtoseconds. The second-harmonic current is generated as electrons drift through the region of momenta corresponding to interband transition resonance at an optical frequency. The resulting strongly asymmetric distortion of carrier distribution in momentum space gives rise to an enhanced electric-dipole nonlinear response at the second harmonic. We develop an approximate analytic theory of this effect which accurately predicts observed intensity and polarization of the second-harmonic signal. 
	
	\end{abstract}
	
	%\pacs{insert here}
	
	\maketitle

\section{Introduction}

There has been a surge of interest in the nonlinear and quantum optics of graphene \cite{bonaccorso2010graphene,kumar2013,hong2013,gu2012regenerative,sun2010,otsuji2012,yao2012,tokman2013,tokman2016,cheng2016,cheng2014,cheng2017second,an2013,an2014,avetissian2012,entin2010,glazov2014,glazov2011,hendry2010,mikhailov2008,mikhailov2016,mikhailov2007,mikhailov2011,mikhailov2014-2,yao2014,constant2016all,wang2016,wang2015,smirnova2014,winzer2013,li2012,konig-otto2016,dean2009,dean2010,lin2014,brun2015,bykov2012,wu2012}. High speed of carriers in monolayer graphene, $v _{F} \approx 10^{8}\ \mathrm{cm/s}$ , and the resulting large dipole moment of the optical transition $\sim ev_F/\omega$, give rise to high magnitudes of the nonlinear optical susceptibilities. A particularly intriguing question raised in recent studies is whether monolayer graphene can support second-order nonlinear processes such as second harmonic or difference frequency generation and other three-wave mixing processes. Since graphene has an in-plane inversion symmetry, the second-order nonlinear response is forbidden in the electric dipole approximation. Of course graphene, like any surface, has anisotropy between in-plane and out-of-plane electron motion, but the corresponding nonlinear response is very weak \cite{dean2009,dean2010}. A much stronger second-order nonlinearity originates from nonlocal response beyond the electric dipole approximation, i.e. magnetic-dipole and electric-quadrupole response \cite{tokman2016,cheng2017second,glazov2014,glazov2011,mikhailov2011,yao2014,constant2016all,wang2016,smirnova2014}. In \cite{lin2014} second-harmonic generation (SHG) was observed in suspended graphene and attributed to the inversion symmetry breaking due to wrinkles and tears in a monolayer. In a bilayer graphene the inversion symmetry can be broken by applying a voltage bias in transverse direction \cite{brun2015}.
 
The most obvious method of creating in-plane anisotropy is the anisotropic perturbation of carrier distribution in the k-space by a constant electric field \cite{cheng2014,an2013,an2014,glazov2014,bykov2012,wu2012}. This can be also achieved by applying a low-frequency field, in particular at THz frequencies. This SHG mechanism corresponds effectively to a third-order nonlinearity. However, since the THz frequency is much lower than the optical frequencies, it is natural to introduce an effective second-order nonlinear response which depends on the THz field amplitude $E_0(t)$ as a parameter \cite{cheng2014}. For a weak low-frequency field the perturbation of an originally isotropic distribution is localized mainly near the Fermi surface. In this case the parameter of anisotropy is expressed through a low-frequency Drude current $J_{Dr}$. The resulting nonlinear response obtained by perturbations turns out to be proportional to the magnitude of $J_{Dr}$ (see \cite{cheng2014}). The magnitude of the SH current $J_{2\omega }$ scales as
\begin{equation} 
\label{Eq:J_SH_weak} 
J_{2\omega} \propto J_{Dr} E_{\omega}^2 ,  
\end{equation} 
where $J_{Dr}=\frac{e^2 \tau_{Dr} \left|\mu \right|}{4\hbar} E_0$, $\tau_{Dr}$ is the relaxation time of a DC current, $\mu$ is chemical potential, $E_{\omega}$ is the electric field amplitude of an optical pumping at frequency $\omega$. Equation \eqref{Eq:J_SH_weak} is valid when the perturbation of electron distribution at Fermi surface is small.

The experimental results that we obtained below apparently agree with the scaling of Eq.~\eqref{Eq:J_SH_weak}. This is however very surprising for ultra-strong fields used in our experiment. Indeed, in recent experiments including our experiment the THz pulses of several hundreds fs duration and field amplitudes $E_0\approx 100 - 300 \ \mathrm{kV/cm}$ were used; see e.g. \cite{tani2012,oladyshkin2017}. Such fields distort the original Fermi distribution far beyond small perturbations: an electron acceleration to energies of the order of 0.2 eV (which is a typical Fermi energy for CVD graphene on a glass substrate) happens over a timescale of only a few fs \cite{oladyshkin2017}. In such strong fields,  a standard expression for the current $J_{Dr}$ using the Drude relaxation time of $\sim 1$ ps adopted in \cite{cheng2014} would yield a current amplitude much higher than  the maximum possible value $e v_F N$, where $N$ is the surface density of carriers! Furthermore, in ultra-strong fields the carrier density gets multiplied by a large factor during the THz pulse due to the electron-hole pair creation \cite{tani2012,oladyshkin2017}. 

Therefore, to interpret our experiments we need a theory which is not restricted to a standard perturbative model of a weak deformation of the Fermi surface. Within our analytic model the anisotropic deformation of the particle distribution is formed primarily in the vicinity of the interband transition resonance between the optical pump and particle states dressed by a low-frequency field. We consider the case when the energies of resonant particles are much higher than the Fermi energy. A dramatic change in the properties of a quantum system dressed by a strong field is a universal effect; see e.g. \cite{kocharovskaya1999} and references therein. In our case this effect shows up as a broadening of the interband resonance due to particle acceleration by a strong low-frequency field $E_0$ in the process of an interband transition intiated by an optical field. The broadening of the resonant region in the k-space, ${\delta k}_s$, corresponds to the frequency bandwidth $\delta \omega = v_F {\delta k}_s \sim {\delta t}_s^{-1}=\sqrt{\frac{e v_F \left|E_0\right|}{\hbar}}$, which is equal to the inverse time of Schwinger pair creation in graphene by the field of magnitude $E_0$ \cite{oladyshkin2017,vajna2015,lewkowicz2011}. Under the action of an ultra-strong field $E_0$ the region of resonant perturbation of carriers in the k-space turns out to be asymmetric with respect to the resonant frequency given by $\omega = 2 k v_F $. This asymmetry gives rise to an anisotropic electric-dipole nonlinear response. The case of an ultrafast electron scattering with characteristic time shorter than the Schwinger time is treated numerically in the Appendix.

Within our model the typical lifetime of the particle perturbation in the resonant region is determined by drift in a low frequency field: $\hbar {\delta k}_s / e E_0 \sim {\delta t}_S$, i.e. it corresponds to the Schwinger time ${\delta t}_S$. Therefore, an increase in the magnitude of a THz field should broaden the applicability region of the perturbation theory with respect to the \textit{optical} field. For the above parameters of the THz pulse the time ${\delta t}_S$ is of the order of $\mathrm{5-8}\ \mathrm{fs}$, which is smaller than, or of the same order as scattering times in graphene. In this case the saturation of the resonant absorption of the optical pump is weak up to the field amplitudes of order $E_{\omega} \sim 2 - 3\ \mathrm{MV/cm}$. This fact ensures the validity of the scaling $J_{2\omega} \propto E^2_{\omega}$ despite the use of high-power femtosecond lasers. 

Section 2 describes the experiment. Section 3 contains the basic set of equations describing SHG. Section 4 describes an approximate analytical model. In Section 5 the theory is compared with experiment. Derivation of certain formulas used in the analytic theory and numerical simulations for ultrafast scattering times can be found in the Appendix.

%%%%%%%%%%%%%%%%%%%%%%%%%%%%%

\section{Experiment}

\begin{figure}[htb]
	\begin{center}
		\includegraphics[scale=0.4]{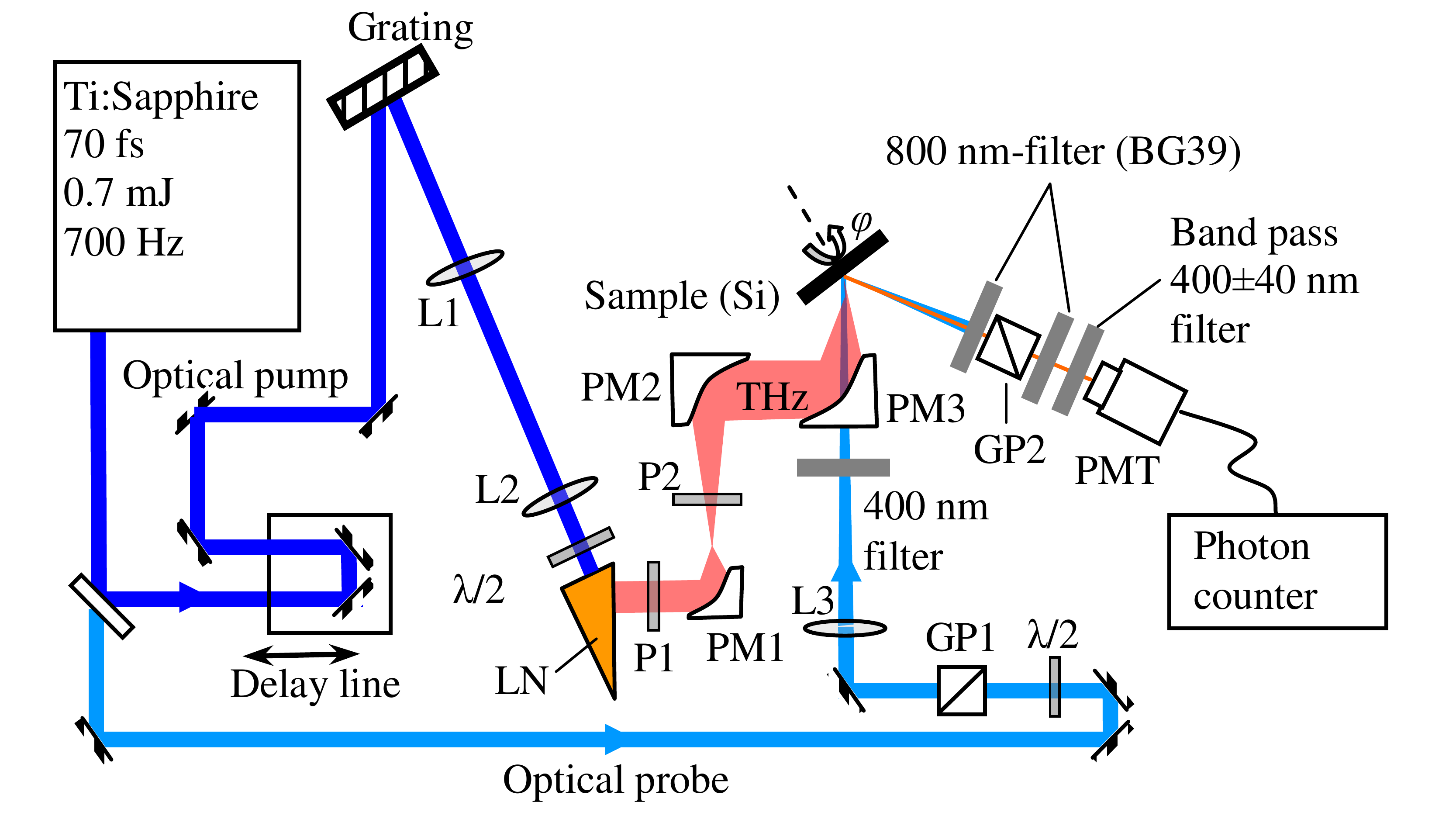}
		\caption{Experimental setup. }
		\label{Fig:exp_setup}
	\end{center}
\end{figure}

The schematic of the experimental setup is shown in Fig.~\ref{Fig:exp_setup}. A Ti-Sapphire laser system (Spitfire, Spectra-Physics) generated pulses of energy 0.7 mJ, central wavelength of 0.795 nm, duration 70 fs and repetition rate 700 Hz. Optical radiation was injected through mirror PM3 parallel to the THz field. The diameter of the optical beam was $\approx 220$ $\mu$m FWHM, the intensity of the optical beam on the sample was 3 $\mathrm{GW/cm^2}$.

THz pulses were generated in the LiNbO${}_{3}$ crystal as in \cite{fulop2010} and focused on the sample at an angle of ${45}^{\circ}$. The diameter of the THz beam on the sample was $\approx 500$ $\mu$m  FWHM with respect to the field amplitude. The maximum THz electric field amplitude was 250 kV/cm. The maximum value of the P-polarized field was two times smaller than that of the S-polarized field.  

The sample was a CVD graphene monolayer on borosilicate glass. Interaction with substrate led to p-doping to the level of Fermi energy $E_F = 0.2$ eV \cite{tani2012}, which is much smaller than the particle energy of 0.75 eV corresponding to the interband resonance with pumping.  

\begin{figure}[htb]
	\begin{center}
		\includegraphics[scale=0.4]{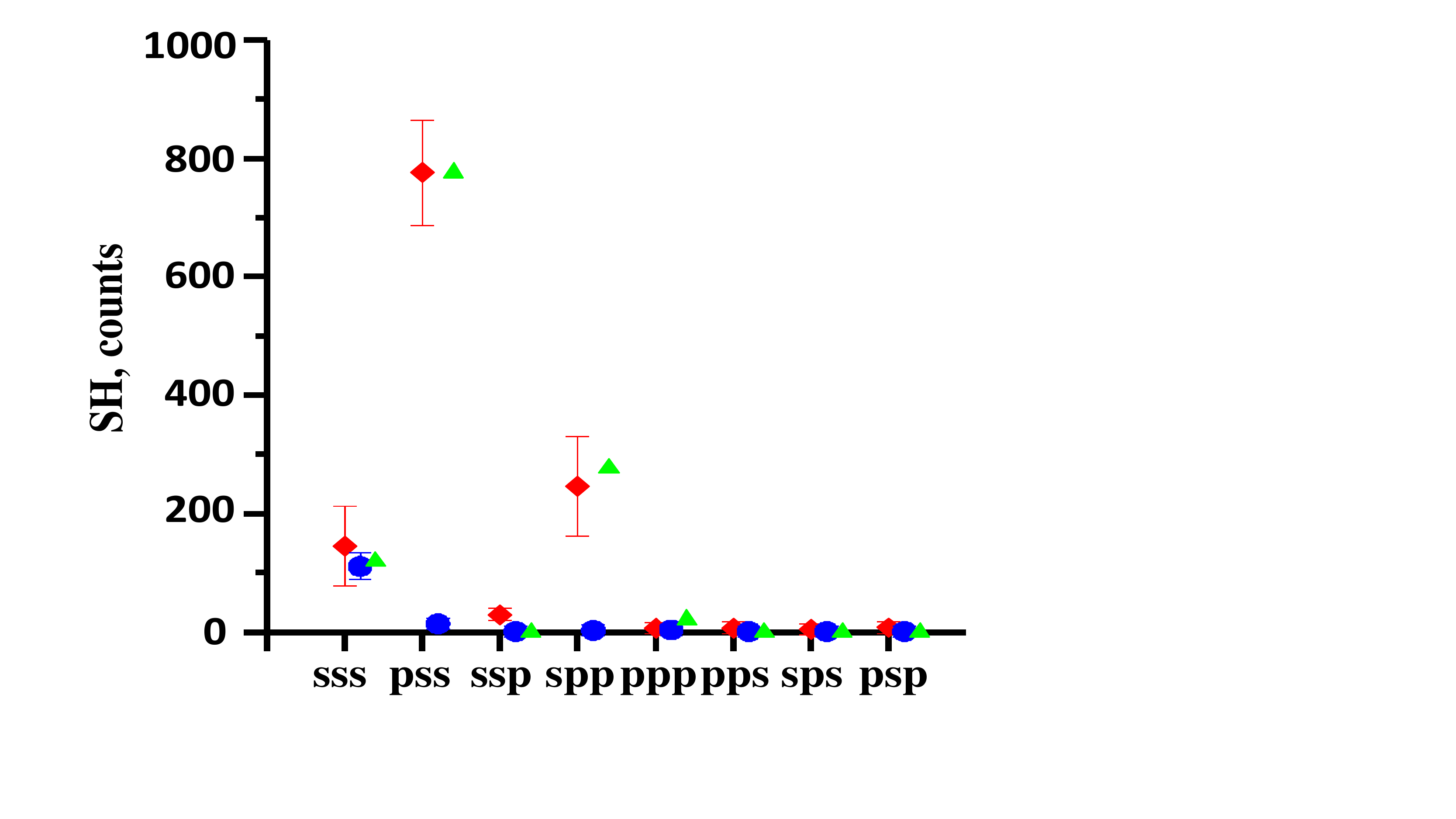}
		\caption{SH signal for different polarizations. Red diamonds are experimental data for graphene on glass, blue circles are for glass without graphene, green triangles is the theory normalized to the experimental signal for SH in pss polarization. }
		\label{Fig:SH_signal}
	\end{center}
\end{figure}

The SH signal for graphene on glass and for glass substrate only is shown in Fig.~\ref{Fig:SH_signal} together with theoretical results. Here p polarization corresponds to the field in the incidence plane, whereas s polarization is orthogonal to this plane. In the notations \textbf{sss}, \textbf{pss} etc. the first index is the polarization of the optical pump, the second index is the polarization of the THz field and the third index is the polarization of SH photons. 	
Clearly, significant SH signal existed only for SH photons with the same polarization as the THz field, in agreement with theory. 

\begin{figure}[htb]
	\begin{center}
		\includegraphics[scale=0.4]{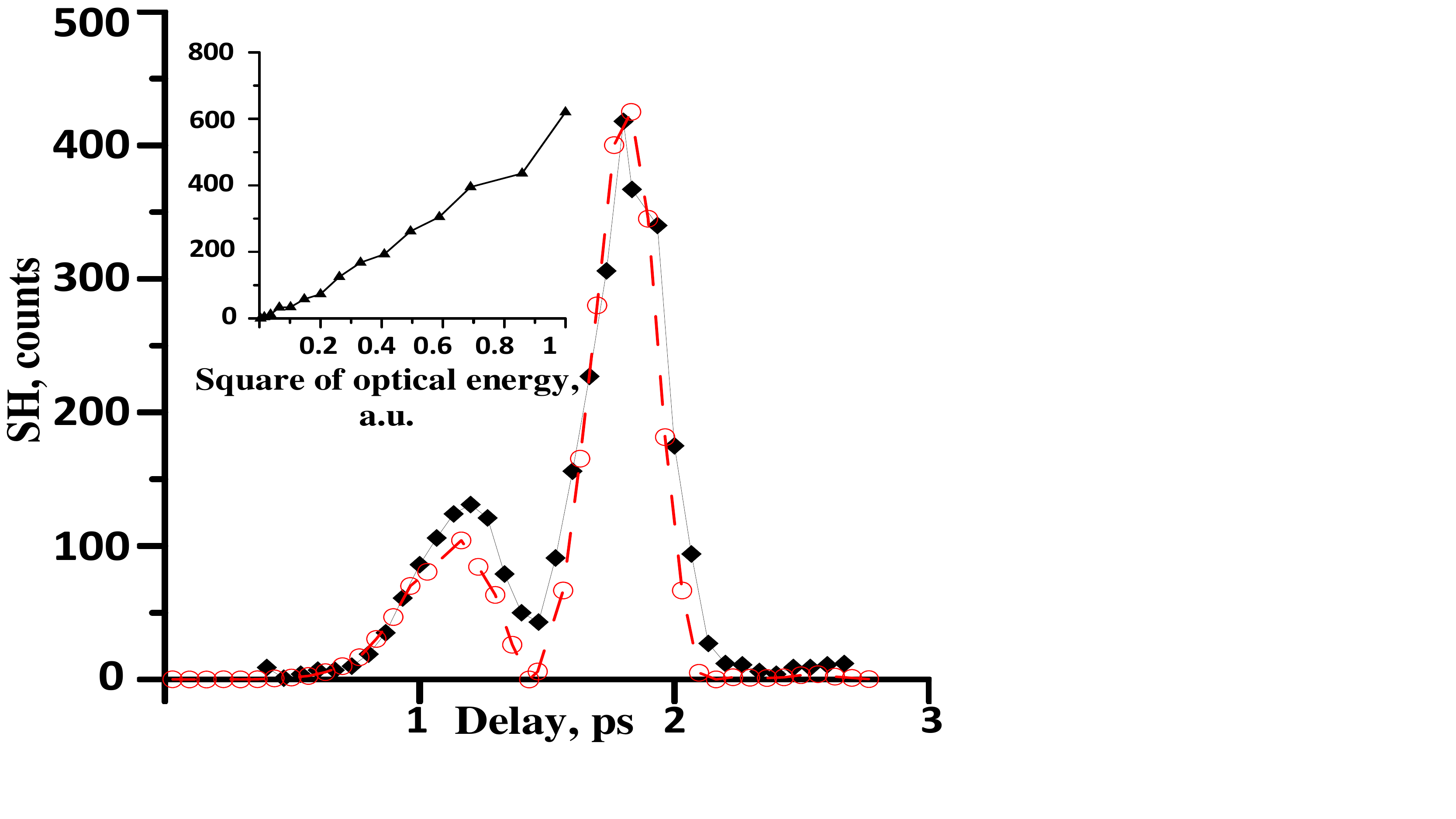}
		\caption{Time dependence of the SH signal and the THz field squared. Black diamonds is the SH signal from the graphene surface, red circles is the THz field squared, normalized to the maximum SH signal. Inset: the dependence of the SH signal from the optical pulse energy squared. }
		\label{Fig:SH_delay}
	\end{center}
\end{figure}

Fig.~\ref{Fig:SH_delay} shows the number of pss-polarized SH photons $N_{2\omega}$ as a function of the delay time between the fs laser pulse and THz pulse, superimposed on the temporal profile of the THz pulse. Clearly, the SHG signal follows the THz field squared, $N_{2\omega} \propto E_0^2$. Since the number of SH photons is proportional to the current $j_{2\omega}$ squared, this corresponds to the scaling Eq.~\eqref{Eq:J_SH_weak}. Inset to Fig. 3 shows the dependence of $N_{2\omega}$ from the energy of a laser pulse, which also agrees with  Eq.~\eqref{Eq:J_SH_weak}. 

Note that the SH radiation propagated in the direction of mirror reflection of the incident optical pump, which proves the coherent nature of the SH signal. 

%%%%%%%%%%%%%%%%%%%%%%%%%%%%%%%

\section{Theoretical model}

\subsection{Equations for the density matrix and current}

Consider monolayer graphene located in the (\textit{x},\textit{y}) plane. Not too far from the Dirac point, the effective Hamiltonian for carriers interacting with an electromagnetic field is \cite{cheng2017second,glazov2014,wang2016}:
\begin{equation} 
\label{GrindEQ__2_} 
\hat{H} = v_{F} \hat{{\mb \sigma}} \cdot \left( \hat{{\mb p}} + {e {\mb A} / c} \right) - e \phi ,                                                                  
\end{equation} 
where $\hat{{\mb p}} = -i\hbar \left({\mb x}_{0} \frac{\partial}{\partial x} + {\mb y}_{0} \frac{\partial}{\partial y} \right)$, $\hat{{\mb \sigma }}={\mb x}_{0} \hat{\sigma }_{x} +{\mb y}_{0} \hat{\sigma }_{y}$, ${\sigma}_{x,y}$ are Pauli matrices, ${\mb{x}}_0,{\mb{y}}_0$ are unit vectors, ${\mb A}$ and $\phi$ are vector and scalar potentials of an external EM field. The basis functions of the Hamiltonian Eq.~\eqref{GrindEQ__3_}, defined at $\boldsymbol{A}\boldsymbol{=}\varphi =0$, are
\begin{equation} \label{GrindEQ__3_} 
{\left| {\mb k},s \right\rangle} =\frac{e^{i{\mb k}\cdot {\mb r}} }{\sqrt{2} } \left(\begin{array}{l} {s} \\ {e^{i\theta_{{\mb k}} } } \end{array} \right),\ s=\pm 1.                                                       
\end{equation} 
Here eigenstates ${\left| {\mb k},s \right\rangle}$ are determined on a unit area for periodic boundary conditions, indices $s=\pm 1$ numerate electron and hole states, $\theta_{{\mb k}}$ is an angle between the quasimomentum $\hbar {\mb k}$ and $x$-axis. The eigenenergy corresponding to states in Eq.~\eqref{GrindEQ__3_} is given by $E_{s} = s\hbar v_{F} \sqrt{k_{x}^{2} + k_{y}^{2}}$ \cite{katsnelson2012graphene}.

In the electric dipole approximation an external electric field can be considered uniform, ${\mb E}(t)$. It can be defined through either vector or scalar potential. The gauge invariance of observables calculated by solving the Schr{\"o}dinger or master equation for the Hamiltonian \eqref{GrindEQ__2_} has been proven in \cite{wang2016}. Therefore, any EM gauge can be used. However, using a scalar potential leads to a slightly simpler derivation when solving a density matrix equation; see the comparison in \cite{wang2016}. Furthermore, according to \cite{tokman2009,zhang2014}, gauge invariance requires that the relaxation operator in the density matrix equation be made dependent on the vector potential, which would be an additional complication. Therefore, we define an external field through the scalar potential, assuming $\phi = - {\mb r}\cdot{\mb E}(t)$, ${\mb A} = 0$ in Eq.~\eqref{GrindEQ__2_}. 

In the absence of relaxation (which we will take into account later) the Von Neumann equation in the basis of Eq.~\eqref{GrindEQ__3_} is
\begin{equation} 
\label{GrindEQ__4_} 
{\dot{\rho}}_{ss'\mb{k}{\mb{k}}'}+i{\upsilon }_F\left(sk-s'k'\right){\rho }_{ss'\mb{k}{\mb{k}}'}=-\frac{i}{\hbar }\sum_{{\mb{s}}''{\mb{k}}''}{\left(V_{ss^{''}\mb{k}{\mb{k}}''}{\rho }_{s^{''}s'{\mb{k}}''{\mb{k}}'}-{\rho }_{ss^{''}\mb{k}{\mb{k}}''}V_{s^{''}s'{\mb{k}}''{\mb{k}}'}\right)},    
\end{equation} 
where
\begin{align} 
\label{GrindEQ__5_} 
&V_{ss'\mb{k}{\mb{k}}'}=\left\langle \mb{k},s\mathrel{\left|\vphantom{\mb{k},s -e\phi  {\mb{k}}',s'}\right.\kern-\nulldelimiterspace}-e\phi \mathrel{\left|\vphantom{\mb{k},s -e\phi  {\mb{k}}',s'}\right.\kern-\nulldelimiterspace}{\mb{k}}',s'\right\rangle =\frac{e}{2}\sum_{x,y}{\left[-\frac{\partial {\theta }_{\mb{k}}}{\partial k_{x,y}}{\delta }_{\mb{k}{\mb{k}}'}+i\left(1+ss'\right)\frac{\partial {\delta }_{\mb{k}{\mb{k}}'}}{\partial k_{x,y}}\right]E_{x,y}\left(t\right)},   \\           
&\frac{\partial {\delta }_{\mb{k}{\mb{k}}'}}{\partial k_{x,y}}=-\frac{\partial {\delta }_{\mb{k}{\mb{k}}'}}{\partial k'_{x,y}},   \ \ 
\frac{\partial {\theta }_{\mb{k}}}{\partial k_x}=-\frac{\sin{\theta }_{\mb{k}}}{k},  \ \  \frac{\partial {\theta }_{\mb{k}}}{\partial k_y}=+\frac{\cos{\theta }_{\mb{k}}}{k},  \ \  k=\sqrt{k_x^2+k_y^2}. \nonumber
\end{align} 

For a uniform field $\mb{E}(t)$ the matrix element of the interaction Hamiltonian $V_{ss'\mb{k}{\mb{k}}'}$ is diagonal with respect to $\mb{k}$, ${\mb{k}}'$. Switching to a continuous \textit{k}-spectrum, Eqs. \eqref{GrindEQ__4_} and \eqref{GrindEQ__5_} yield a closed set of equations for matrix elements ${\rho}_{ss'\mb{k}\mb{k}}$. Denoting the quantum coherence as ${\rho }_{+1-1\mb{k}\mb{k}}={\rho }_{\mb{k}}$, population difference as ${\rho }_{-1-1\mb{k}\mb{k}}-{\rho }_{+1+1\mb{k}\mb{k}}={\mathrm{\Delta }}_{\mb{k}}$, and the sum of populations as ${\rho }_{+1+1\mb{k}\mb{k}}+{\rho }_{-1-1\mb{k}\mb{k}}=n_{\mathrm{\Sigma}\mb{k}}$, we obtain
\begin{equation} 
\label{GrindEQ__6_} 
\left(\frac{\partial }{\partial t}+i{\omega }_k-\frac{e\mb{E}}{\hbar }\frac{\partial }{\partial \mb{k}}\right){\rho }_{\mb{k}}=-i\frac{{\mathrm{\Omega }}_{\mb{k}}}{2}{\mathrm{\Delta }}_{\mb{k}},                                                                 
\end{equation} 
\begin{equation} \label{GrindEQ__7_} 
\left(\frac{\partial }{\partial t}-\frac{e\mb{E}}{\hbar }\frac{\partial }{\partial \mb{k}}\right){\mathrm{\Delta }}_{\mb{k}}=-i{\mathrm{\Omega }}_{\mb{k}}\left({\rho }_{\mb{k}}-{\rho }^*_{\mb{k}}\right),                                                                
\end{equation} 
\begin{equation} \label{GrindEQ__8_} 
\left(\frac{\partial }{\partial t}-\frac{e\mb{E}}{\hbar }\frac{\partial }{\partial \mb{k}}\right)n_{\mathrm{\Sigma }\mb{k}}=0,                                                                                     
\end{equation} 
where $\omega_{k} =2v_{F} k$ is the interband transition frequency, ${\mathrm{\Omega }}_{\mb{k}}=\frac{e}{k\hbar }\left(\sin{\theta }_{\mb{k}}E_x\mb{-}\cos{\theta }_{\mb{k}}E_y\right)$ the Rabi frequency. Equation \eqref{GrindEQ__8_} is separated from the rest of the system and we won't use it anymore.

Eqs. \eqref{GrindEQ__6_} and \eqref{GrindEQ__7_} make a closed set of equations, which is a version of semiconductor Bloch equations \cite{cheng2014,alnaib2015}. Note that the ``convective'' terms $\sim \frac{\partial }{\partial \mb{k}}$ in Eqs. \eqref{GrindEQ__6_}-\eqref{GrindEQ__8_} do not originate from some kind of phenomenological assumptions, e.g. an attempt to make it look like a Boltzmann-type equation or quasiclassical equations of motion with a Berry field. These terms rigorously follow from the matrix elements of the interaction Hamiltonian Eq.~\eqref{GrindEQ__5_} in a given gauge. If we defined the same field $\mb{E}(t)$ through the vector potential, we would get the equations in a different form, see e.g.~\cite{malic2011}. This would not change the observables of course.  

The current operator is defined by $\hat{\mb{J}}=-(i e / \hbar) \left[\hat{H},\mb{r}\right]=-ev_F\hat{\mb{\sigma }}$. The observable current is $\mb{J} = \sum_{s,s',\mb{k}=\mb{k}'}{{\mb{j}}_{ss'\mb{k}\mb{k}'} \rho_{s's{\mb{k}}'\mb{k}}}$, where ${\mb{j}}_{ss'\mb{k}{\mb{k}}'}$ are matrix elements of the current operator. Using ${\rho }_{ss'\mb{k}{\mb{k}}'}\propto {\delta }_{\mb{k}{\mb{k}}'}$  and performing integration in \textit{k}-space, we obtain 
\begin{equation} \label{GrindEQ__9_} 
j_x=ev_F\frac{g}{4{\pi }^2}\int_{\infty }{d^2k\left[\cos{\theta }_{\mb{k}}{\mathrm{\Delta} }_{\mb{k}}+i\sin{\theta }_{\mb{k}}\left({\rho }_{\mb{k}}-{\rho }^*_{\mb{k}}\right)\right.},                                                      
\end{equation} 
\begin{equation} \label{GrindEQ__10_} 
j_y=ev_F\frac{g}{4{\pi }^2}\int_{\infty }{d^2k\left[\sin{\theta }_{\mb{k}}{\mathrm{\Delta} }_{\mb{k}}-i\cos{\theta }_{\mb{k}}\left({\rho }_{\mb{k}}-{\rho }^*_{\mb{k}}\right)\right.},                                                    
\end{equation} 
where g = 4 is spin and valley degeneracy. The first term in square brackets in Eqs.~\eqref{GrindEQ__9_},\eqref{GrindEQ__10_} is the intraband current and the second term is an interband current. 

%%%%%%%%%%%%%%%%%%%%%%%%%%%%

\subsection{Equations for slow variables}

Now we make the following ansatz which separates the slow THz dynamics (subscript 0) from fast optical frequencies:
\[\mb{E}(t) = E_0(t)+\mathrm{Re}{\mb{E}}_{\omega }(t){\mathrm{e}}^{-i\omega t}\] 
\[{\mathrm{\Omega }}_{\mb{k}}(t) = {\mathrm{\Omega }}_{0\mb{k}}(t)+\mathrm{Re}{\mathrm{\Omega }}_{\omega \mb{k}}(t){\mathrm{e}}^{-i\omega t}\] 
\[\mathrm{\Delta }_{\mb{k}} = \mathrm{\Delta }_{0\mb{k}}(t)+\mathrm{Re}{\mathrm{\Delta} }_{\omega \mb{k}}(t){\mathrm{e}}^{-i\omega t}+\mathrm{Re}{\mathrm{\Delta} }_{2\omega \mb{k}}(t){\mathrm{e}}^{-i2\omega t}\] 
\[{\rho }_{\mb{k}}={\rho }_{0\mb{k}}(t)+{\rho }^{\left(-\right)}_{\omega \mb{k}}(t){\mathrm{e}}^{-i\omega t}+{\rho }^{\left(+\right)}_{\omega \mb{k}}(t){\mathrm{e}}^{+i\omega t}+{\rho }^{\left(-\right)}_{2\omega \mb{k}}(t){\mathrm{e}}^{-2i\omega t}+{\rho }^{\left(+\right)}_{2\omega \mb{k}}(t){\mathrm{e}}^{+i2\omega t}\] 
It also introduces envelopes of the optical fields at the fundamental and second harmonics.

In addition, we add the relaxation operator in its simplest form to the right-hand side of  Eqs. \eqref{GrindEQ__6_}, \eqref{GrindEQ__7_}. This leads to the replacement  ${\omega }_k\Longrightarrow {\omega }_k-i\gamma $ in Eq. \eqref{GrindEQ__6_} and the population relaxation term $-\mathrm{\Gamma }\left({\mathrm{\Delta }}_{\mb{k}}-{\mathrm{\Delta }}_{kF}\right)$ in Eq. \eqref{GrindEQ__7_}. Here $\gamma ,\ \mathrm{\Gamma }$ are relaxation rates  of the coherence and population difference and  ${\mathrm{\Delta }}_{kF}$ is the equilibrium Fermi distribution. 

The resulting equations below contain only these slow-varying envelopes. However, we still keep counter-rotating terms, as shown below.
\begin{equation} 
\label{GrindEQ__11_} 
\left(\frac{\partial }{\partial t}+\gamma +i{\omega }_k-\frac{e{\mb{E}}_0}{\hbar }\frac{\partial }{\partial \mb{k}}\right){\rho }_{0\mb{k}}=-\frac{i}{2}{\mathrm{\Omega }}_{0\mb{k}}{\mathrm{\Delta }}_{0\mb{k}}-\frac{i}{4}\mathrm{Re}\left({\mathrm{\Omega }}^*_{\omega \mb{k}}{\mathrm{\Delta }}_{\omega \mb{k}}\right)+\frac{e}{2\hbar }\left({\mb{E}}^{\mb{*}}_{\omega }\frac{\partial {\rho }^{\left(-\right)}_{\omega \mb{k}}}{\partial \mb{k}}+e{\mb{E}}_{\omega }\frac{\partial {\rho }^{\left(+\right)}_{\omega \mb{k}}}{\partial \mb{k}}\right) 
\end{equation} 
\begin{equation} \label{GrindEQ__12_} 
\left(\frac{\partial }{\partial t}+\gamma +i{\omega }_k-i\omega -\frac{e{\mb{E}}_0}{\hbar }\frac{\partial }{\partial \mb{k}}\right){\rho }^{\left(-\right)}_{\omega \mb{k}}=-\frac{i}{4}\left({\mathrm{\Omega }}_{\omega \mb{k}}{\mathrm{\Delta }}_{0\mb{k}}+{\mathrm{\Omega }}_{0\mb{k}}{\mathrm{\Delta }}_{\omega \mb{k}}\right)+\frac{1}{2}\frac{e{\mb{E}}_{\omega }}{\hbar }\frac{\partial {\rho }_{0\mb{k}}}{\partial \mb{k}} 
\end{equation} 
\begin{equation} \label{GrindEQ__13_} 
\left(\frac{\partial }{\partial t}+\gamma +i{\omega }_k+i\omega -\frac{e{\mb{E}}_0}{\hbar }\frac{\partial }{\partial \mb{k}}\right){\rho }^{\left(+\right)}_{\omega \mb{k}}=-\frac{i}{4}\left({\mathrm{\Omega }}^*_{\omega \mb{k}}{\mathrm{\Delta }}_{0\mb{k}}+{\mathrm{\Omega }}_{0\mb{k}}{\mathrm{\Delta }}^*_{\omega \mb{k}}\right)+\frac{1}{2}\frac{e{\mb{E}}^{\mb{*}}_{\omega }}{\hbar }\frac{{\partial \rho }_{0\mb{k}}}{\partial \mb{k}} 
\end{equation} 
\[\left(\frac{\partial }{\partial t}-\frac{e{\mb{E}}_0}{\hbar }\frac{\partial }{\partial \mb{k}}\right){\mathrm{\Delta }}_{0\mb{k}}+\mathrm{\Gamma }\left({\mathrm{\Delta }}_{0\mb{k}}-{\mathrm{\Delta }}_{Fk}\right)=-i{\mathrm{\Omega }}_{0\mb{k}}\left({\rho }_{0\mb{k}}-{\rho }^*_{0\mb{k}}\right)+\frac{e}{2\hbar }\mathrm{Re}\left({\mb{E}}^{\mb{*}}_{\omega }\frac{\partial {\Delta }_{\omega \mb{k}}}{\partial \mb{k}}\right)\mb{+}\] 
\begin{equation} \label{GrindEQ__14_} 
\mathrm{+}\mathrm{Im}\left({\mathrm{\Omega }}^*_{\omega \mb{k}}{\rho }^{\left(-\right)}_{\omega \mb{k}}+{\mathrm{\Omega }}_{\omega \mb{k}}{\rho }^{\left(+\right)}_{\omega \mb{k}}\right) 
\end{equation} 
\begin{equation} \label{GrindEQ__15_} 
\left(\frac{\partial }{\partial t}-i\omega -\frac{e{\mb{E}}_0}{\hbar }\frac{\partial }{\partial \mb{k}}\right){\mathrm{\Delta }}_{\omega \mb{k}}+\mathrm{\Gamma }{\mathrm{\Delta }}_{\omega \mb{k}}=-i{\mathrm{\Omega }}_{\omega \mb{k}}\left({\rho }_{0\mb{k}}-{\rho }^*_{0\mb{k}}\right)-i2{\mathrm{\Omega }}_{0\mb{k}}\left({\rho }^{\left(-\right)}_{\omega \mb{k}}-{\rho }^{\left(+\right)*}_{\omega \mb{k}}\right)+\frac{e}{\hbar }{\mb{E}}_{\omega }\frac{\partial {\mathrm{\Delta }}_{0\mb{k}}}{\partial \mb{k}} 
\end{equation} 
In Eqs. \eqref{GrindEQ__11_}-\eqref{GrindEQ__15_} the contribution of the terms oscillating at 2$\omega$ is neglected since the SH field is very weak. The perturbation at the second harmonics is described by
\begin{equation} \label{GrindEQ__16_} 
\left(\frac{\partial }{\partial t}+\gamma +i{\omega }_k-i2\omega -\frac{e{\mb{E}}_0}{\hbar }\frac{\partial }{\partial \mb{k}}\right){\rho }^{\left(-\right)}_{2\omega \mb{k}}=\frac{1}{2}\frac{e{\mb{E}}_{\omega }}{\hbar }\frac{\partial {\rho }^{\left(-\right)}_{\omega \mb{k}}}{\partial \mb{k}}-\frac{i}{8}{\mathrm{\Omega }}_{\omega \mb{k}}{\mathrm{\Delta }}_{\omega \mb{k}} 
\end{equation} 
\begin{equation} \label{GrindEQ__17_} 
\left(\frac{\partial }{\partial t}+\gamma +i{\omega }_k+i2\omega -\frac{e{\mb{E}}_0}{\hbar }\frac{\partial }{\partial \mb{k}}\right){\rho }^{\left(+\right)}_{2\omega \mb{k}}=\frac{1}{2}\frac{e{\mb{E}}^{\mb{*}}_{\omega }}{\hbar }\frac{{\partial \rho }^{\left(+\right)}_{\omega \mb{k}}}{\partial \mb{k}}-\frac{i}{8}{\mathrm{\Omega }}^*_{\omega \mb{k}}{\mathrm{\Delta }}^*_{\omega \mb{k}} 
\end{equation} 
\begin{equation} \label{GrindEQ__18_} 
\left(\frac{\partial }{\partial t}-i2\omega -\frac{e{\mb{E}}_0}{\hbar }\frac{\partial }{\partial \mb{k}}\right){\mathrm{\Delta }}_{2\omega \mb{k}}+\mathrm{\Gamma }{\mathrm{\Delta }}_{2\omega \mb{k}}=-i{\mathrm{\Omega }}_{\omega \mb{k}}\left({\rho }^{\left(-\right)}_{\omega \mb{k}}-{\rho }^{\left(+\right)*}_{\omega \mb{k}}\right)+\frac{e}{2\hbar }{\mb{E}}_{\omega }\frac{\partial {\mathrm{\Delta }}_{\omega \mb{k}}}{\partial \mb{k}} 
\end{equation} 
In numerical modeling the solution to Eqs.~\eqref{GrindEQ__11_} - \eqref{GrindEQ__15_} was substituted into Eqs.~\eqref{GrindEQ__16_} - \eqref{GrindEQ__18_}, whereas the solution to Eqs.~\eqref{GrindEQ__16_} - \eqref{GrindEQ__18_} was substituted into Eqs.~\eqref{GrindEQ__9_} - \eqref{GrindEQ__10_} to find the SH current.

%%%%%%%%%%%%%%%%%%%%%%%%%%%%%

\section{An approximate analytic solution}

\subsection{Equations in rotating wave approximation}

Within the rotating wave approximation (RWA) we can solve for the dynamics of carriers in the vicinity of an interband transition at the fundamental frequency of the optical field. The RWA corresponds to the following inequalities:
\[ \left|{\omega }_k-\omega \right|, \ \gamma, \ \mathrm{\Gamma }\ll \omega \  \textrm{and} \  \left|\frac{e{\mb{E}}_0}{\hbar }\frac{\partial }{\partial \mb{k}}\right|\sim \frac{1}{{\delta t}_s}\ll \omega . \]
In this case one can neglect counter-rotating terms ${\rho }^{\left(+\right)}_{\omega \mb{k}}{\mathrm{e}}^{+i\omega t}$ and ${\rho }^{\left(+\right)}_{2\omega \mb{k}}{\mathrm{e}}^{+2i\omega t}$ in terms  of \cref{GrindEQ__11_,GrindEQ__12_,GrindEQ__13_,GrindEQ__14_,GrindEQ__15_,GrindEQ__16_,GrindEQ__17_,GrindEQ__18_}. Furthermore, if the following conditions are satisfied,
\begin{equation} \label{GrindEQ__19_} 
\left|{\mathrm{\Omega }}_{0\mb{k}}\right|,\ \left|{\mathrm{\Omega }}_{\omega \mb{k}}\right|\ll \left(\gamma +\frac{1}{{\delta t}_s}\right),\ \left(\mathrm{\Gamma }+\frac{1}{{\delta t}_s}\right)\ll \omega. 
\end{equation} 
One can assume that $\left|{\mathrm{\Delta} }_{0\mb{k}}\right|\gg \left|{\mathrm{\Delta} }_{\omega \mb{k}}\right|$, $\left|{\rho }_{0\mb{k}}\right|\ll \left|{\rho }^{\left(-\right)}_{\omega \mb{k}}\right|$ in \cref{GrindEQ__11_,GrindEQ__12_,GrindEQ__13_,GrindEQ__14_,GrindEQ__15_,GrindEQ__16_,GrindEQ__17_,GrindEQ__18_} and obtain approximate equations, 
\begin{equation} \label{GrindEQ__20_} 
\left(\frac{\partial }{\partial t}+\gamma +i{\omega }_k-i\omega -\frac{e{\mb{E}}_0}{\hbar }\frac{\partial }{\partial \mb{k}}\right){\rho }_{\omega \mb{k}}=-\frac{i}{4}{\mathrm{\Omega }}_{\omega \mb{k}}{\mathrm{\Delta }}_{0\mb{k}}, 
\end{equation} 
\begin{equation} \label{GrindEQ__21_} 
\left(\frac{\partial }{\partial t}-\frac{e{\mb{E}}_0}{\hbar }\frac{\partial }{\partial \mb{k}}\right){\mathrm{\Delta }}_{0\mb{k}}+\mathrm{\Gamma }\left({\mathrm{\Delta }}_{0\mb{k}}-{\mathrm{\Delta }}_{Fk}\right)=\mathrm{Im}\left({\mathrm{\Omega }}^*_{\omega \mb{k}}{\rho }_{\omega \mb{k}}\right),                                           
\end{equation} 
\begin{equation} \label{GrindEQ__22_} 
{\rho }_{2\omega \mb{k}}\approx \frac{i}{\omega }\frac{e{\mb{E}}_{\omega }}{2\hbar }\frac{\partial {\rho }_{\omega \mb{k}}}{\partial \mb{k}},   {\Delta }_{2\omega \mb{k}}\approx \frac{{\mathrm{\Omega }}_{\omega \mb{k}}}{2\omega }{\rho }_{\omega \mb{k}},                                                                
\end{equation} 
where ${\rho }_{\mb{k}}\approx {\rho }_{0\mb{k}}+{\rho }_{\omega \mb{k}}{\mathrm{e}}^{-i\omega t}+{\rho }_{2\omega \mb{k}}{\mathrm{e}}^{-2i\omega t}$.

%%%%%%%%%%%%%%%%%%%%%%%%%%%

\subsection{The stationary phase solution of density matrix equations }

For simplicity we consider the case when optical and THz fields are polarized along \textit{x}; other polarizations are considered in Sec.~IV D. Using ${\mathrm{\Omega }}_{\omega \mb{k}}(t)=\frac{e{\hbar }^{-1}\sin{\theta }_{\mb{k}}E_{\omega }(t)}{k}=\frac{k_ye{\hbar }^{-1}E_{\omega }(t)}{k^2_y+k^2_x}$ and $\frac{e{\mb{E}}_0(t)}{\hbar }\frac{\partial }{\partial \mb{k}}=\frac{eE_0(t)}{\hbar }\frac{\partial }{\partial k_x}$, the formal solution of Eq.~\eqref{GrindEQ__20_} can be written as
\begin{equation} \label{GrindEQ__23_} 
{\rho }_{\omega \mb{k}}(t)=-\frac{i}{4}\int^t_{-t_0}{{\mathrm{e}}^{i{\mathrm{\Psi }}_{\mb{k}}\left(t',t\right)-\gamma \cdot \left({t-t}'\right)}{\mathrm{\Omega }}_{\omega \mb{k}}\left(t',t\right){\mathrm{\Delta }}_{0\mb{k}}\left(t',t\right)dt'},                       
\end{equation} 
where
\begin{equation} \label{GrindEQ__24_} 
{\mathrm{\Psi }}_{\mb{k}}\left(t',t\right)=-\int^t_{t'}{{\omega }_{\mb{k}}\left(\tau ,t\right)d\tau +\omega \cdot \left({t-t}'\right)}.                                                    
\end{equation} 
The functions ${\mathrm{\Omega }}_{\omega \mb{k}}\left(t',t\right)$, ${\mathrm{\Delta }}_{0\mb{k}}\left(t',t\right)$ and ${\omega }_{\mb{k}}\left(\tau ,t\right)$ are obtained from functions  ${\mathrm{\Omega }}_{\omega \mb{k}}(t)$, ${\mathrm{\Delta }}_{0\mb{k}}(t)$ and ${\omega }_k=2{\upsilon }_F\sqrt{k^2_y+k^2_x}$ as
\[{\mathrm{\Omega }}_{\omega \mb{k}}\left(t',t\right) = \frac{k_ye{\hbar }^{-1}E_{\omega }\left(t'\right)}{k^2_y+{\left(k_x+\frac{e}{\hbar }\int^t_{t'}{E_0\left(\tau \right)d\tau }\right)}^2},\] 
\[{\mathrm{\Delta }}_{0\mb{k}}\left(t',t\right) = {\mathrm{\Delta }}_{0\mb{k}}\left(t',k_y,k_x+\frac{e}{\hbar }\int^t_{t'}{E_0\left(\tau \right)d\tau }\right),\] 
\[{\omega }_{\mb{k}}\left(\tau ,t\right) = 2v_F\sqrt{k^2_y+{\left(k_x+\frac{e}{\hbar }\int^t_{\tau }{E_0\left({\tau }'\right)d{\tau }'}\right)}^2}.\] 

It is easy to check that \cref{GrindEQ__23_,GrindEQ__24_} is an exact solution to Eq.~\eqref{GrindEQ__20_}, satisfying the initial condition ${\rho }_{\omega \mb{k}}\left(-t_0\right)=0$, where $t=-t_0$ is the time when the optical field is turned on. 

The integral in Eq.~\eqref{GrindEQ__23_} can be evaluated by the method of stationary phase. Within this approach one assumes that the main contribution to the integral comes from the vicinity of a stationary point given by the condition  
\begin{equation} \label{GrindEQ__25_} 
{\left.\frac{\mathrm{\partial }{\mathrm{\Psi }}_{\mb{k}}\left(t',t\right)}{\partial t'}\right|}_{t'=t_s}={\left.{\omega }_{\mb{k}}\left(t',t\right)\right|}_{t'=t_s}-\omega =0.                                                                 
\end{equation} 
Expanding ${\mathrm{\Psi }}_{\mb{k}}\left(t',t\right)$ near the stationary point $t'=t_s\left(t,\mb{k}\right)$, we obtain
\begin{equation} \label{GrindEQ__26_} 
{\rho }_{\omega \mb{k}}(t)\approx -\frac{i}{4}{\mathrm{e}}^{i{\mathrm{\Psi }}_{\mb{k}}\left(t_s,t\right)-\gamma \left(t-t_s\right)}{\mathrm{\Omega }}_{\omega \mb{k}}\left(t_s,t\right){\mathrm{\Delta }}_{0\mb{k}}\left(t_s,t\right)\int^{t-t_s}_{-\left(t_0+t_s\right)}{{\mathrm{e}}^{i{\alpha }_{\mb{k}}\left(t_s,t\right){\tau }^2}d\tau }.                 
\end{equation} 
where ${\mathrm{\Psi }}_{\mb{k}}\left(t_s,t\right)={\left.{\mathrm{\Psi }}_{\mb{k}}\left(t',t\right)\right|}_{t'=t_s}$,
\begin{equation} \label{GrindEQ__27_} 
{\alpha }_{\mb{k}}\left(t_s,t\right)=\frac{1}{2}{\left.\frac{{\partial }^2{\mathrm{\Psi }}_{\mb{k}}\left(t',t\right)}{\partial t^{'2}}\right|}_{t'=t_s}=\frac{1}{2}{\left.\frac{\partial {\omega }_{\mb{k}}\left(t',t\right)}{\partial t'}\right|}_{t'=t_s}.                                       
\end{equation} 

The interval  in Eq.~\eqref{GrindEQ__26_} which makes the main contribution to the integral is given by
\[\mathrm{\Delta} \tau =\sqrt{{1}/{\left|{\alpha }_{\mb{k}}\right|}}{\sim \delta t}_S=\sqrt{\frac{\hbar }{ev_F\left|E_0\right|}},\] 
The following hierarchy of timescales ensures the validity of the stationary phase method:
\begin{subequations}
\begin{align}
{\delta t}_S\ll {\mathrm{\Delta} t}_{opt},\ \ {\mathrm{\Delta} t}_{THz},\ \ {\gamma }^{-1} , \label{GrindEQ__28a_} 
\intertext{where ${\mathrm{\Delta t}}_{opt}$ and ${\mathrm{\Delta} t}_{THz}$ are durations of the optical and THz pulses. Furthermore, we assume that } 
{\mathrm{\Delta} t}_{opt}\ll {\mathrm{\Delta} t}_{THz},\ \   {\delta t}_s\omega \gg 1,  \label{GrindEQ__28b_}
\end{align}
\end{subequations}
Eq.~\eqref{GrindEQ__26_} can be written in the form (see Appendix A) 
\begin{align} \label{GrindEQ__29_} 
{\rho }_{\omega \mb{k}}=-\frac{i}{4}{\mathrm{\Omega }}_{\omega \mb{k}}\left(t_s,t\right){\mathrm{\Delta }}_{0\mb{k}}\left(t_s,t\right)\frac{\pi }{2v_F}\mathfrak{W}\left(Z,{\delta k}_s\right),                                           
\end{align} 
where
\begin{equation} \label{GrindEQ__30_} 
\mathfrak{W}\left(Z,{\delta k}_s\right)=\frac{{\mathrm{e}}^{\pm iZ^2\ }}{{\delta k}_s}\left(\frac{{\mathrm{e}}^{\mp i{\pi }/{4}}}{\sqrt{\pi }}\mp \frac{2}{\pi }\int^Z_0{{\mathrm{e}}^{\mp ix^2}dx}\right),                                              
\end{equation} 
$Z=\frac{k-\frac{\omega }{2v_F}}{{\delta k}_s}$, ${\delta k}_s=\sqrt{\left|\cos{\theta }_{\mb{k}}\frac{eE_0}{\hbar v_F}\right|}$. The top and bottom signs in Eq.~\eqref{GrindEQ__30_} correspond to $\cos{\theta }_{\mb{k}}E_0 > 0$ and $\cos{\theta }_{\mb{k}}E_0 < 0$ respectively. 

The function $\mathfrak{W}$ is normalized as $\int^{\infty }_{-\infty }{\mathfrak{W}\left(Z,{\delta k}_s\right){\delta k}_sdZ}=1$. In the limit ${\delta k}_s\rightarrow 0$ (i.e. when $E_0\rightarrow 0$) Eqs.~\eqref{GrindEQ__29_}, \eqref{GrindEQ__30_} give
\[{\rho }_{\omega \mb{k}}\approx \frac{1}{4}{\mathrm{\Omega }}_{\omega \mb{k}}{\mathrm{\Delta }}_{0\mb{k}}\left[\frac{\mathcal{P}}{{\omega -\omega }_k\ }-i\pi \delta \left({\omega -\omega }_k\mp \epsilon \right)\right],\] 
where $\mathcal{P}$ indicates a principal value of the integral, $\epsilon \rightarrow +0$.

One can see from the expression in Eq.~\eqref{GrindEQ__30_} that the scattering-induced broadening of the resonance $\delta \omega \sim \gamma $ is replaced by the nonlinear field-induced broadening: $\delta \omega \sim {1}/{{\delta t}_s\propto \sqrt{\left|E_0\right|}}$. The absorption line described by Eq.~\eqref{GrindEQ__30_} is asymmetric, namely it is shifted towards ${\omega >\omega }_k$ for $\cos{\theta }_{\mb{k}}E_0>0$ and towards ${\omega <\omega }_k$ for $\cos{\theta }_{\mb{k}}E_0<0$. An asymmetric shape is due to the drift of carriers in the k-space in the presence of a THz field.  

The functions ${\mathrm{\Omega }}_{\omega \mb{k}}\left(t_s,t\right)$ and ${\mathrm{\Delta }}_{0\mb{k}}\left(t_s,t\right)$ in Eq.~\eqref{GrindEQ__29_} are Lagrangian variables ${\mathrm{\Omega }}_{\omega \mb{k}}\left(\mb{k}(t),t\right)$ and ${\mathrm{\ }\mathrm{\Delta }}_{0\mb{k}}\left(\mb{k}(t),t\right)$, determined at the current moment of time \textit{t} shifted with respect to a stationary point $t_s$ given by Eq.~\eqref{GrindEQ__25_}. The shift amount is $t-t_s\approx \mp \frac{Z}{{\delta k}_sv_F}\sim Z{\delta t}_s$, see Appendix. The possibility to use a ``local'' approximation ${\mathrm{\Omega }}_{\omega \mb{k}}\left(t_s,t\right)\approx {\mathrm{\Omega }}_{\omega \mb{k}}(t)$ and ${\mathrm{\Delta }}_{0\mb{k}}\left(t_s,t\right)\approx {\mathrm{\Delta }}_{0\mb{k}}(t)$ is discussed in Appendix A.  It turns out that the first and the second inequality in Eqs.~(28a) always ensure the validity the local approximation for the Rabi frequency. At the same time, the local approximation for the population difference ${\mathrm{\Delta }}_{0\mb{k}}(t)$ requires in addition that the perturbation of the initial value ${\mathrm{\Delta }}_{0\mb{k}}={\mathrm{\Delta} }_{Fk}$ in the resonant region be small enough. The perturbation of the population difference can be estimated using Eqs. \eqref{GrindEQ__21_}, \eqref{GrindEQ__29_}. The analysis in Appendix A yields 
\begin{equation} \label{GrindEQ__31_} 
\left|{\mathrm{\Delta }}_{0\mb{k}}-{\mathrm{\Delta} }_{Fk}\right|\sim {\delta t}^2_s\frac{{\left|{\mathrm{\Omega }}_{\omega \mb{k}}\right|}^2}{4}.                                                     
\end{equation} 

Equation \eqref{GrindEQ__31_} looks like a standard result for a two-level system coupled to an optical field if we replace the population relaxation time with ${\delta t}_S$ in the density matrix equations. This is to be expected, since the lifetime of the perturbation in the presence of a strong low-frequency field is determined by the drift of carriers out of the resonant region: ${\hbar {\delta k}_s}/{eE_0}\sim {\delta t}_S$. Using ${\delta t}^2_s\frac{{\left|{\mathrm{\Omega }}_{\omega \mb{k}}\right|}^2}{4}\approx \frac{{\left|E_{\omega }\right|}^2}{{\left|E_0\right|}^2}\frac{1}{{\delta t}_s^2{\omega }^2}$ , $\omega \sim 2.5\ {\mathrm{fs}}^{-1}$ and ${\delta t}_s$ longer than 5 fs (i.e. for $E_0$ smaller than 300 kV/cm) one can obtain $\left|{\mathrm{\Delta }}_{0\mb{k}}-1\right|\approx 0.3$ for $\frac{{\left|E_{\omega }\right|}^2}{{\left|E_0\right|}^2}\approx 45$. Here we also assumed ${\mathrm{\Delta} }_{Fk}=1$, which is the case for high enough electron energies corresponding to the interband resonance ${\omega }_k\approx \omega $.

%%%%%%%%%%%%%%%%%%%%%%%%

\subsection{The nonlinear current}

For the electric field polarized along \textit{x,} Eqs. \eqref{GrindEQ__9_} and \eqref{GrindEQ__22_} give the following expression for the complex amplitude of the current at SH: 
\begin{equation} \label{GrindEQ__32_} 
j_{x2\omega }=-\frac{e v_F g}{8\omega {\pi }^2}\int_{\infty }{\left({\mathrm{\Omega }}_{\omega \mb{k}}{\rho }_{\omega \mb{k}}\cos{\theta }_{\mb{k}}\right)d^2k}.                                                        
\end{equation} 
Substituting Eq.~\eqref{GrindEQ__29_} into Eq.~\eqref{GrindEQ__32_} and assuming ${\mathrm{\Delta }}_{0\mb{k}}\left(t_s,t\right)\approx 1$ and ${\mathrm{\Omega }}_{\omega \mb{k}}\left(t_s,t\right)\approx {\mathrm{\Omega }}_{\omega \mb{k}}(t)$ leads to
\begin{equation} \label{GrindEQ__33_} 
j_{x2\omega }\approx i\frac{ge^3E^2_{\omega }}{16\pi {\hbar }^2\omega }\int^{\pi }_{-\pi }{d{\theta }_{\mb{k}}\int^{\infty }_0{\frac{dk}{k}\mathfrak{W}\left(Z,{\delta k}_s\right){\sin}^2{\theta }_{\mb{k}}\cos{\theta }_{\mb{k}}}}\ ,                                    
\end{equation} 
where $\mathfrak{M}\left(Z,{\delta k}_s\right)$ is defined in Eq.~\eqref{GrindEQ__30_}. Limiting ourselves to the resonant region, we can expand $\frac{1}{k}\approx \frac{1}{k_{res}}-\frac{{\delta k}_sZ}{k^2_{res}}+\frac{{\delta k}^2_sZ^2}{{2k}^3_{res}}\dots $ when evaluating the inegral in Eq. \eqref{GrindEQ__33_} (here $k_{res}=\frac{\omega }{2v_F}$) and change from $\int^{\infty }_0{dk\dots }$ to infinite integration limits when integrating over $dZ=\frac{dk\ }{{\delta k}_s}$:
\begin{equation} \label{GrindEQ__34_} 
j_{x2\omega }\approx i\frac{ge^3E^2_{\omega }}{16\pi {\hbar }^2\omega }\int^{\pi }_{-\pi }{d{\theta }_{\mb{k}}\int^{\infty }_{-\infty }{dZ\left(\frac{1}{k_{res}}-\frac{{\delta k}_sZ}{k^2_{res}}+\frac{{\delta k}^2_sZ^2}{{2k}^3_{res}}\dots \right){\delta k}_s\mathfrak{W}\left(Z,{\delta k}_s\right){\sin}^2{\theta }_{\mb{k}}\cos{\theta }_{\mb{k}}}}\ .    
\end{equation} 
The first two terms in the expansion in Eq.~\eqref{GrindEQ__34_} give zero after angle integration $\int^{\pi }_{-\pi }{d{\theta }_{\mb{k}}\dots }$ . For the first term this is obvious due to the normalization condition $\int^{\infty }_{-\infty }{\mathfrak{M}\left(Z,{\delta k}_s\right){\delta k}_sdZ}=1$; the proof for the second term is in the Appendix. To calculate the integral for the third term in the expansion we take into account that the function $\mathfrak{M}$ is a linear combination of the even function ${\mathrm{e}}^{\pm iZ^2}$ and odd function ${\mathrm{e}}^{\pm iZ^2}\int^Z_0{{\mathrm{e}}^{\mp ix^2}dx}$, see Eq.~\eqref{GrindEQ__30_}. As a result, we obtain
\[j_{x2\omega }\approx i\frac{ge^3v^3_FE^2_{\omega }}{4\pi {\hbar }^2{\omega }^4}\int^{\pi }_{-\pi }{d{\theta }_{\mb{k}}{\sin}^2{\theta }_{\mb{k}}\cos{\theta }_{\mb{k}}\left|\cos{\theta }_{\mb{k}}\ \frac{eE_0}{\hbar v_F}\right|\int^{\infty }_{-\infty }{\frac{{\mathrm{e}}^{\pm i\left(Z^2-\ \frac{\pi }{4}\right)\ }}{\sqrt{\pi }}Z^2dZ}}\ ,\] 
where $\int^{\infty }_{-\infty }{\frac{{\mathrm{e}}^{\pm i\left(Z^2-\ \frac{\pi }{4}\right)\ }}{\sqrt{\pi }}Z^2dZ}=\mp i$. Recalling that the upper and lower signs correspond to $\cos{\theta }_{\mb{k}}E_0>0$ and $\cos{\theta }_{\mb{k}}E_0<0$, we get
\begin{equation} \label{GrindEQ__35_} 
j_{x2\omega }\approx \frac{ge^4v^2_FE^2_{\omega }E_0}{4\pi {\hbar }^3{\omega }^4}\int^{\pi }_{-\pi }{{d{\theta }_{\mb{k}}\sin}^2{\theta }_{\mb{k}}{\cos}^2{\theta }_{\mb{k}}}=\frac{e^4v^2_FE^2_{\omega }E_0}{4{\hbar }^3{\omega }^4}.                                             
\end{equation} 

%%%%%%%%%%%%%%%%%%%%%%%%%%%%

\subsection{Polarization dependence}
\label{sect:pol_dep}

Eq.~\eqref{GrindEQ__35_} has been obtained for collinear vectors ${\mb{E}}_{\omega }\parallel {\mb{E}}_0 \parallel {\mb{j}}_{2\omega }$. Consider now a different orientation when ${\mb{E}}_{\omega } \perp {\mb{E}}_0 \parallel {\mb{j}}_{2\omega }$. Let ${\mb{E}}_0 \parallel {\mb{x}}_0$ and ${\mb{j}}_{2\omega }\parallel {\mb{x}}_0$, but ${\mb{E}}_{\omega }\parallel {\mb{y}}_0$. The expression for the Rabi frequency ${\mathrm{\Omega }}_{\omega \mb{k}}$ takes the form ${\mathrm{\Omega }}_{\omega \mb{k}}=\mb{-}\frac{\cos{\theta }_{\mb{k}}}{k}\frac{eE_{\omega }}{\hbar }$. It is straightforward to obtain that the relationship between the SH current $j_{x2\omega }$ and ${\mathrm{\Omega }}_{\omega \mb{k}}$ has the same form as in Eq. \eqref{GrindEQ__32_}. Therefore, instead of Eq.~\eqref{GrindEQ__35_} we obtain
\begin{equation} \label{GrindEQ__36_} 
j_{x2\omega }\approx \frac{ge^4v^2_FE^2_{\omega }E_0}{4\pi {\hbar }^3{\omega }^4}\int^{\pi }_{-\pi }{d{\theta }_{\mb{k}}{\cos}^4{\theta }_{\mb{k}}}=\frac{3e^4v^2_FE^2_{\omega }E_0}{4{\hbar }^3{\omega }^4}.                                               
\end{equation} 
Comparing Eqs.~\eqref{GrindEQ__35_} with \eqref{GrindEQ__36_} one can see that the power of the SHG signal at ${\mb{E}}_{\omega } \perp {\mb{E}}_0$ is 9 times higher than at ${\mb{E}}_{\omega }\parallel {\mb{E}}_0$ for the same magnitudes of the fields.

In the case ${\mb{E}}_0 \perp {\mb{j}}_{2\omega }$ we always get ${\mb{j}}_{2\omega }=0$ for any polarization of the optical field. This is expected, since for this geometry the low-frequency field cannot break the inversion symmetry along the direction of ${\mb{j}}_{2\omega }$ see the Appendix. 

%%%%%%%%%%%%%%%%%%%%%%%%%%%%%%%%%%

\section{Comparison with experiment}

The field dependence and polarization dependence of the observed SH signal coincides with those predicted by the model. To compare absolute numbers of SH photons, one needs to take into account that 

\noindent (\textbf{i}) Eqs. \eqref{GrindEQ__35_}, \eqref{GrindEQ__36_} contain the components of the fields tangential to the monolayer. Using Fresnel formulas \cite{landau1984}, one can get  
\begin{equation} \label{GrindEQ__37_} 
E^{\left(S\right)}_{\tau }=\frac{2E^{\left(S\right)}\cos\theta }{\sqrt{\varepsilon -{\sin}^2\theta }+\cos\theta },\ \ 
E^{\left(P\right)}_{\tau }=\frac{2E^{\left(P\right)}\cos\theta \sqrt{\varepsilon -{\sin}^2\theta }}{\sqrt{\varepsilon -{\sin}^2\theta }+\varepsilon \cos\theta }.                                           
\end{equation} 
Here $E^{\left(S,P\right)}$ is the amplitude of the transverse field of an incident S- or P-polarized wave, $E^{\left(S,P\right)}_{\tau }$ is its tangential component on the monolayer, $\varepsilon $ the dielectric constant of the substrate, $\theta$ the incidence angle from air with respect to the normal to the surface of graphene.  

\noindent (\textbf{ii}) The surface current $j^{\left(S,P\right)}_{2\omega }$  at frequency $2\omega $ and the amplitude $E^{\left(S,P\right)}_{2\omega }$ of the S- or P-polarized field radiated by this current are related by
\begin{equation} \label{GrindEQ__38_} 
E^{\left(S\right)}_{2\omega }=-\frac{4\pi }{c}\frac{j^{\left(S\right)}_{2\omega }}{\sqrt{\varepsilon -{\sin}^2\theta }+\cos\theta }, E^{\left(P\right)}_{2\omega }=-\frac{4\pi }{c}\frac{j^{\left(P\right)}_{2\omega }\sqrt{\varepsilon -{\sin}^2\theta }}{\sqrt{\varepsilon -{\sin}^2\theta }+\varepsilon \cos\theta },                               
\end{equation} 
where \textit{$\theta$ }is the reflection angle.

\noindent (\textbf{iii}) The powers of the S- and P-polarized THz radiation in the experiment differ by a factor of 4. 

For numerical estimates we assume ${\varepsilon }_{\omega }\approx {\varepsilon }_{2\omega }\approx 2.25$, ${\varepsilon }_0\approx 6.25$, $\theta \approx {45}^{{}^\circ }$.

Next we calculate the number of photons in the PSS polarization configuration. From Eqs. \eqref{GrindEQ__36_} $-$ \eqref{GrindEQ__38_} for the above parameters ${\varepsilon }_{\omega }$, ${\varepsilon }_{2\omega }$, ${\varepsilon }_0$ and $\theta$ we can estimate the field of the SH signal as
\begin{equation} \label{GrindEQ__39_} 
E^{\left(S\right)}_{2\omega }\approx 0.3\frac{4\pi }{c}\frac{e^4v^2_F{\left(E^{\left(P\right)}_{\omega }\right)}^2 E^{\left(S\right)}_0} {8{\hbar }^3{\omega }^4},                                                               
\end{equation} 
which is equal to $3.8 \times {10}^{-3}~\mathrm{esu}$. For the pulse duration and beam cross-section used in the experiment, and for a 7\% experimental efficiency of the detection system, the SH field in Eq. \eqref{GrindEQ__39_} corresponds to about 850 SH photons per series of 60,000 laser pulses. This agrees with an experimentally measured SH signal within the experimental accuracy. 

Relative SHG efficiency for other polarizations can be obtained from Eqs.~\eqref{GrindEQ__35_}$-$\eqref{GrindEQ__38_}:
\begin{equation} \label{GrindEQ__40_} 
\frac{N^{PSS}_{2\omega }}{N^{SSS}_{2\omega }}=9{\left[{\alpha}^{\left({P}/{S}\right)}_{\omega }\right]}^2,   \ \
\frac{N^{PSS}_{2\omega }}{N^{SPP}_{2\omega }}=\frac{{4\left[{\alpha}^{\left({P}/{S}\right)}_{\omega }\right]}^2}{{\alpha}^{\left({P}/{S}\right)}_0{\alpha}^{\left({P}/{S}\right)}_{2\omega }}, \ \ 
\frac{N^{PSS}_{2\omega }}{N^{PPP}_{2\omega }}=\frac{36}{{\alpha}^{\left({P}/{S}\right)}_0{\alpha}^{\left({P}/{S}\right)}_{2\omega }},                        
\end{equation} 
where
\[{\alpha}^{\left({P}/{S}\right)}_{\omega ,2\omega ,0}={\left({\varepsilon }_{\omega ,2\omega ,0}-{\sin}^2\theta \right)\left[\frac{\sqrt{{\varepsilon }_{\omega ,2\omega ,0}-{\sin}^2\theta }+\cos\theta }{\sqrt{{\varepsilon }_{\omega ,2\omega ,0}-{\sin}^2\theta }+{\varepsilon }_{\omega ,2\omega ,0}\cos\theta }\right]}^2.\] 
For the same parameters we get ${\alpha}^{\left({P}/{S}\right)}_{\omega ,2\omega }\approx 0,84$, ${\alpha}^{\left({P}/{S}\right)}_0\approx 1.2$, which gives ${N^{PSS}_{2\omega }}/{N^{SSS}_{2\omega }}\approx 6.4$, ${N^{PSS}_{2\omega }}/{N^{SPP}_{2\omega }\approx 2,8}$, ${N^{PSS}_{2\omega }}/{N^{PPP}_{2\omega }}\approx 36$.

These theoretical values are compared with experiment in Fig.~2. As is clear from the figure, there is a good agreement between theory and experiment. 

%%%%%%%%%%%%%%%%%%%%%%%%%%%%%%%%%%%

\begin{acknowledgments} 
This work has been supported by the RFBR grant No.~18-29-19091. Y.W.~and A.B.~acknowledge the support by Air Force Office for Scientific Research
through Grants No.~FA9550-17-1-0341 and FA9550-14-1-0376.

\end{acknowledgments}

%%%%%%%%%%%%%%%%%%%%%%%%%%

\appendix

\section{The stationary phase solution for the density matrix equation}

According to the first inequality in Eq.~(28a) the optical pulse is longer than the integration interval in Eq. \eqref{GrindEQ__26_}. Therefore the lower limit of integration can be taken as $-\infty $; therefore, Eq.~\eqref{GrindEQ__26_} yields
\begin{align} \label{GrindEQ__A1_}
{\rho }_{\omega \mb{k}}(t)\approx -\frac{i}{4}{\mathrm{e}}^{i{\mathrm{\Psi }}_{\mb{k}}\left(t_s,t\right)-\gamma \left(t-t_s\right)}{\mathrm{\Omega }}_{\omega \mb{k}}\left(t_s,t\right){\mathrm{\Delta }}_{0\mb{k}}\left(t_s,t\right)\times \left(\frac{{\mathrm{e}}^{\pm i{\pi }/{4}}}{2}\sqrt{\frac{\pi }{\left|{\alpha }_{\mb{k}}\left(t_s,t\right)\right|}}+\int^{t-t_s}_0{{\mathrm{e}}^{i{\alpha }_{\mb{k}}\left(t_s,t\right)x^2}dx}\right) ,
\end{align}
where the upper and lower signs in ${\mathrm{e}}^{\pm i{\pi }/{4}}$ correspond to ${\alpha }_{\mb{k}}>0$ and ${\alpha }_{\mb{k}}<0$ respectively. Furthermore, from Eqs. \eqref{GrindEQ__25_}, \eqref{GrindEQ__24_} one can get
\begin{equation} \label{GrindEQ__A2_}
{\left.\frac{\mathrm{\partial }{\mathrm{\Psi }}_{\mb{k}}\left(t',t\right)}{\partial t'}\right|}_{t'=t_s}{=\omega }_{\mb{k}}\left(t_s,t\right)-\omega =2v_F\sqrt{k^2_y+{\left(k_x+e{\hbar }^{-1}\int^t_{t_s}{E_0\left(\tau \right)d\tau }\right)}^2}-\omega =0 .
\end{equation}
Since the THz pulse is much longer than the optical pulse, one can write the integral $\int^t_{t_s}{E_0\left(\tau \right)d\tau }$ as 
\begin{equation} \label{GrindEQ__A3_}
\int^t_{t_s}{E_0\left(\tau \right)d\tau }\approx E_0 \cdot \left(t-t_s\right) ,
\end{equation}
where $E_0$ can be treated as a constant during the optical pulse. 

The calculations are greatly simplified if we assume 
\begin{equation} \label{GrindEQ__A4_}
\frac{e}{\hbar }\left|E_0\right|\left(t-t_s\right)\ll k\sim k_{res}=\frac{\omega }{2v_F} ;
\end{equation}
we will check its validity later. 

Substituting Eq. \eqref{GrindEQ__A3_} into Eq. \eqref{GrindEQ__A2_} and using Eq. \eqref{GrindEQ__A4_}, we obtain:
\begin{equation}   \label{GrindEQ__A5_}
t-t_s\approx \frac{\omega -{\omega }_k}{2\cos{\theta }_{\mb{k}}\frac{v_FeE_0}{\hbar }} ;    
\end{equation}
Using Eq. \eqref{GrindEQ__A5_} in Eqs. \eqref{GrindEQ__24_}, \eqref{GrindEQ__27_}, gives
\begin{align}  
&{\mathrm{\Psi }}_{\mb{k}}\left(t_s,t\right) \approx \frac{{\left(\omega -{\omega }_k\right)}^2}{4\cos{\theta }_{\mb{k}}\frac{v_FeE_0}{\hbar }} , \label{GrindEQ__A6_} \\
&{\alpha }_{\mb{k}}\left(t_s,t\right) \approx -\cos{\theta }_{\mb{k}}\frac{v_FeE_0}{\hbar } . \label{GrindEQ__A7_}
\end{align}
Substituting Eqs. \eqref{GrindEQ__A6_}, \eqref{GrindEQ__A7_} into Eq. \eqref{GrindEQ__A1_} results in
\begin{align}  \label{GrindEQ__A8_}
{\rho }_{\omega \mb{k}}(t) &\approx -\frac{i}{4}{\mathrm{e}}^{i\frac{{\left(\omega -{\omega }_k\right)}^2}{4\cos{\theta }_{\mb{k}}\frac{ev_FE_0}{\hbar }} - \gamma \frac{\omega -{\omega }_k}{2\cos{\theta }_{\mb{k}}\frac{ev_FE_0}{\hbar }}}{\mathrm{\Omega }}_{\omega \mb{k}}\left(t_s,t\right){\mathrm{\Delta }}_{0\mb{k}}\left(t_s,t\right) \nonumber \\
&\times \left({\mathrm{e}}^{\mp i{\pi }/{4}}\sqrt{\frac{\pi }{\left|\cos{\theta }_{\mb{k}}\frac{ev_F E_0}{\hbar }\right|}}+\int^{\left(\frac{\omega -{\omega }_k}{2\cos{\theta }_{\mb{k}}\frac{v_F eE_0}{\hbar }}\right)}_0{{\mathrm{e}}^{-i\cos{\theta }_{\mb{k}}\frac{ev_F E_0}{\hbar }{\tau }^2}d\tau }\right) ,
\end{align}
where the upper and lower signs in ${\mathrm{e}}^{\mp i{\pi }/{4}}$ are for $\cos{\theta }_{\mb{k}}E_0>0$ and $\cos{\theta }_{\mb{k}}E_0<0$ respectively. Taking into account that the integration interval which makes the main contribution near the resonance is $\left|\omega -{\omega }_k\right|{\delta t}_s\sim 1$ and using \cref{GrindEQ__A5_}, we obtain
\[
\frac{e{\hbar }^{-1}\left|E_0\right|\left(t-t_s\right)}{k_{res}}\sim \frac{1}{{\delta t}_s2v_Fk_{res}}\sim \frac{1}{{\delta t}_s\omega },
\] 
which means that the last inequality in \cref{GrindEQ__28b_} ensures the validity of the approximation \cref{GrindEQ__A4_}. In the region $\left|\omega -{\omega }_k\right|{\delta t}_s\sim 1$ the factor $\mathrm{exp}\left(-\ \gamma \frac{\omega -{\omega }_k}{2v_F\cos{\theta }_{\mb{k}}e{\hbar }^{-1}E_0}\right)$ in \cref{GrindEQ__A8_} cannot be greater than $\mathrm{exp}\left(-\ \gamma {\delta t}_s\right)$, so if the last inequality in \cref{GrindEQ__28a_} is satisfied, it can be taken as 1. As a result, we obtain Eqs. \eqref{GrindEQ__29_}, \eqref{GrindEQ__30_}. We also provide here useful asymptotics of the function $\mathfrak{W}\left(Z,{\delta k}_s\right)$ in Eq. \eqref{GrindEQ__30_} at $\left|Z\right|\gg 1$:
\begin{align} \label{GrindEQ__A9_}
\mathfrak{M}\left(Z,{\delta k}_s\right)\approx \frac{1}{i\pi {\delta k}_s}\frac{1}{Z}+\left\{ \begin{array}{lr}
0 & \mbox{\ \ \ for $Z E_0 \cos{\theta }_{\mb{k}}>0$} \\ 
\frac{2{\mathrm{e}}^{\pm i\left(Z^2-\frac{\pi }{4}\right)}}{\sqrt{\pi }{\delta k}_s} & \mbox{\ \ \ for $Z E_0 \cos{\theta }_{\mb{k}}<0$} \end{array}
\right. ,
\end{align}
where the upper and lower signs ${\mathrm{e}}^{\pm i\left(Z^2-\frac{\pi }{4}\right)}$ are taken for $\cos{\theta }_{\mb{k}}E_0>0$ and $\cos{\theta }_{\mb{k}}E_0<0$ respectively.

\section{Approximate expressions for ${{\mathrm{\Omega }}}_{\omega \mb{k}}\left(t_s,t\right)$ and ${\mathrm{\Delta}}_{0\mb{k}}\left(t_s,t\right)$ }

The functions ${\mathrm{\Omega }}_{\omega \mb{k}}\left(t_s,t\right){\mathrm{,\ \ }\mathrm{\Delta }}_{0\mb{k}}\left(t_s,t\right)$ can be written for the same conditions \cref{GrindEQ__A4_} and \cref{GrindEQ__A5_} as
\begin{align}
{\mathrm{\Omega }}_{\omega \mb{k}}\left(t_s,t\right) 
&\approx {\mathrm{\Omega }}_{\omega \mb{k}}(t)-\frac{\omega -{\omega }_k}{2v_F\cos{\theta }_{\mb{k}}\frac{eE_0}{\hbar }}\left(\frac{\partial }{\partial t}-\frac{eE_0}{\hbar }\frac{\partial }{\partial k_x}\right){\mathrm{\Omega }}_{\omega \mb{k}}  \nonumber \\
&=\frac{\sin{\theta }_{\mb{k}}}{k}\frac{e}{\hbar }\left(1-\frac{\omega -{\omega }_k}{2v_F\cos{\theta }_{\mb{k}}\frac{eE_0}{\hbar }}\frac{\partial }{\partial t}-\frac{\omega -{\omega }_k}{{kv}_F}\right)E_{\omega }(t)       \label{GrindEQ__B1_}  \\                             %(A2.1)
{\mathrm{\Delta }}_{0\mb{k}}\left(t_s,t\right)
&\approx {\mathrm{\Delta }}_{0\mb{k}}(t)-\frac{\omega -{\omega }_k}{2v_F\cos{\theta }_{\mb{k}}\frac{eE_0}{\hbar }}\left(\frac{\partial }{\partial t}-\frac{eE_0}{\hbar }\frac{\partial }{\partial k_x}\right){\mathrm{\Delta }}_{0\mb{k}}(t)  \label{GrindEQ__B2_}                               %(A2.2)
\end{align}
 
Taking into account that $\left|\frac{\omega -{\omega }_k}{2v_F\cos{\theta }_{\mb{k}}\frac{eE_0}{\hbar }}\right|\sim {\delta t}_s$, and ${\delta t}_s\ll {\mathrm{\Delta} t}_{opt}$, we can drop the terms with time derivative $\frac{\partial }{\partial t}$ in \cref{GrindEQ__B1_} and \cref{GrindEQ__B2_}. The factor $\frac{\omega -{\omega }_k}{{kv}_F}\sim \frac{{\delta k}_s}{k}$ is small too. At the same time the term $\frac{eE_0}{\hbar }\frac{\omega -{\omega }_k}{2v_F\cos{\theta }_{\mb{k}}\frac{eE_0}{\hbar }}\frac{\partial {\mathrm{\Delta }}_{0\mb{k}}(t)}{\partial k_x}\sim \frac{\partial {\mathrm{\Delta }}_{0\mb{k}}(t)}{\partial k_x}\cdot \frac{k\ -\ \frac{\omega }{2v_F}}{\cos{\theta }_{\mb{k}}}\sim \frac{\left|{\mathrm{\Delta }}_{0\mb{k}}-{\Delta }_{FK}\right|}{{\delta k}_s}\frac{k\ -\ \frac{\omega }{2v_F}}{\cos{\theta }_{\mb{k}}}$ in \cref{GrindEQ__B2_} is small only if the perturbation of the population difference in the resonant region is small too. Therefore, we obtain 
\begin{align} 
&{\mathrm{\Omega }}_{\omega \mb{k}}\left(t_s,t\right)\approx {\mathrm{\Omega }}_{\omega \mb{k}}(t)         \label{GrindEQ__B3_} \\      % (A2.3)
&{\mathrm{\Delta }}_{0\mb{k}}\left(t_s,t\right)\approx {\mathrm{\Delta }}_{0\mb{k}}(t)-\frac{\partial {\mathrm{\Delta }}_{0\mb{k}}(t)}{\partial k_x}\cdot \frac{k\ -\ \frac{\omega }{2v_F}}{\cos{\theta }_{\mb{k}}}        \label{GrindEQ__B4_}             % (A2.4)
\end{align}

\section{Quasistationary perturbation of populations}

At high carrier energies ${\omega }_k\approx \omega $ one can take ${\mathrm{\Delta} }_{Fk}=1$ in Eq. \eqref{GrindEQ__21_}. Consider a stationary solution of this equation:   
\begin{equation} \label{GrindEQ__C1_}
{\mathrm{\Delta} }_{0\mb{k}}-1=-\frac{\hbar }{eE_0}{\mathrm{e}}^{\frac{\hbar \mathrm{\Gamma }}{eE_0}k_x}\int^{k_x}_C{{\mathrm{e}}^{-\frac{\hbar \mathrm{\Gamma }}{eE_0}k'_x}\mathrm{Im}\left[{\mathrm{\Omega }}^*_{\omega {\mb{k}}'}{\rho }_{\omega {\mb{k}}'}\right]dk'_x} ,                      % (A3.1)
\end{equation}
where the notation ${\mb{k}}'$ means ${\mb{k}}'=\ \mb{k}\left({k'_x,k}_y\right)$. The boundary \textit{C} of the integration limit in \cref{GrindEQ__C1_} is chosen where the effective source $\mathrm{Im}\left[{\mathrm{\Omega }}^\ast_{\omega {\mb{k}}'}{\rho }_{\omega {\mb{k}}'}\right]$ approaches zero. This choice depends also on the sign of $E_{0}$, which determines the direction of particle drift in the k-space: for $E_{0} >0$ we get $\left(k_x-k'_x\right)<0$, whereas for $E_{0} <0$ we get $\left(k_x-k'_x\right)>0$. Next, we substitute expressions Eqs. \eqref{GrindEQ__29_}, \eqref{GrindEQ__30_} into \cref{GrindEQ__C1_} and take into account that the characteristic ``size'' of the source $\mathrm{Im}\left[{\mathrm{\Omega }}^\ast_{\omega {\mb{k}}'}{\rho }_{\omega {\mb{k}}'}\right]$ in the \textit{k}-space ${\delta k}_s\sim {1}/{{\delta t}_sv_F}$. As a result, under the condition $\frac{\hbar \mathrm{\Gamma }}{eE_0}{\delta k}_s\ll 1$ we arrive at Eq.\eqref{GrindEQ__31_} for the perturbation of populations. Note that the inequality $\frac{\hbar \mathrm{\Gamma }}{eE_0}{\delta k}_s\ll 1$ is equivalent to ${\delta t}_s\mathrm{\Gamma }\mathrm{\ll }\mathrm{1}$.

\section{The second term in the nonlinear current expansion}

Here we prove that the second term in the expansion in Eq.\eqref{GrindEQ__34_} is equal to zero. Its expression is given by
\[I_2=A \cdot {\mathrm{lim}}_{\epsilon \longrightarrow 0}\int^{\pi }_{-\pi }{d{\theta }_{\mb{k}}\int^{\infty }_{-\infty }{dZ{\mathrm{e}}^{-\epsilon Z^2\ }Z{\delta k}^2_s\mathfrak{W}\left(Z,{\delta k}_s\right){\sin}^2{\theta }_{\mb{k}}\cos{\theta }_{\mb{k}}}}\ ,\] 
where $\epsilon >0$ and $A$ is a constant. We introduced the factor ${\mathrm{e}}^{-\epsilon Z^2\ }$ which makes the proof easier, and we will take the limit $\epsilon =0$ in the end. Taking into account that the function $\mathfrak{M}$ is a linear combination of the even function ${\mathrm{e}}^{\pm iZ^2}$ and odd function ${\mathrm{e}}^{\pm iZ^2}\int^Z_0{{\mathrm{e}}^{\mp ix^2}dx}$ (see Eq.~\eqref{GrindEQ__30_}), we obtain
\[I_2=A \cdot {\mathrm{lim}}_{\epsilon \longrightarrow 0}\int^{\pi }_{-\pi }{{d{\theta }_{\mb{k}}\sin}^2{\theta }_{\mb{k}}\cos{\theta }_{\mb{k}}\sqrt{\left|\cos{\theta }_{\mb{k}}\right|}Y}\ ,\] 
where
\[Y=\mp \int^{\infty }_{-\infty }{Z{\mathrm{e}}^{\left(\pm i-\epsilon \right)Z^2 }\left(\int^Z_0{{\mathrm{e}}^{\mp ix^2}dx}\right)dZ}=\frac{\sqrt{{\pi }/{\epsilon }}}{2\left(i\mp \epsilon \right)}.\] 
In the last expression upper and lower signs are given by the signs of $\cos{\theta }_{\mb{k}}E_0$, as usual. Therefore we get 
\[I_2=A\cdot {\mathrm{lim}}_{\epsilon \longrightarrow 0}\sqrt{\frac{\pi }{\epsilon }}\cdot \frac{\epsilon }{{\epsilon }^2+1}\int^{{\pi }/{2}}_{-{\pi }/{2}}{{d{\theta }_{\mb{k}}\sin}^2{\theta }_{\mb{k}}{\cos}^{\frac{3}{2}}{\theta }_{\mb{k}}}=0.\]

\section{Polarization selection rules for THz field-induced SHG}

Consider the orientation ${\mb{E}}_{\omega }\parallel {\mb{y}}_0$, ${\mb{E}}_0\parallel {\mb{y}}_0$ and ${\mb{j}}_{2\omega }\parallel {\mb{x}}_0$. In this case one should take ${\mathrm{\Omega }}_{\omega \mb{k}}=\mb{-}\frac{\cos{\theta }_{\mb{k}}}{k}\frac{eE_{\omega }}{\hbar }$, $\frac{e{\mb{E}}_0(t)}{\hbar }\frac{\partial }{\partial \mb{k}}=\frac{eE_0(t)}{\hbar }\frac{\partial }{\partial k_y}$ and ${\rho }_{2\omega \mb{k}}\approx \frac{i}{\omega }\frac{eE_{\omega }}{2\hbar }\frac{\partial {\rho }_{\omega \mb{k}}}{\partial k_y}$ in Eqs. \eqref{GrindEQ__20_}$-$\eqref{GrindEQ__22_}. The dependence of $j_{x2\omega }$ on ${\mathrm{\Omega }}_{\omega \mb{k}}$ has the same form as in Eq.~\eqref{GrindEQ__31_}, whereas in Eq.~\eqref{GrindEQ__29_} one has to replace ${\delta k}_s=\sqrt{\left|\cos{\theta }_{\mb{k}} \ \frac{eE_0}{\hbar v_F}\right|}$ with ${\delta k}_s=\sqrt{\left|\sin{\theta }_{\mb{k}} \ \frac{eE_0}{\hbar v_F}\right|}$. Upper and lower signs in all coefficients in Eq. \eqref{GrindEQ__29_} correspond to the signs of $\sin{\theta }_{\mb{k}}E_0$. As a result, instead of Eq.\eqref{GrindEQ__35_} we get
\[j_{x2\omega }\approx \frac{g e^4v^2_FE^2_{\omega }E_0}{8\pi {\hbar }^3{\omega }^4}\int^{\pi }_{-\pi }{d{\theta }_{\mb{k}}\sin{\theta }_{\mb{k}}{\cos}^3{\theta }_{\mb{k}}}=0.\] 

If ${\mb{E}}_{\omega }\parallel {\mb{x}}_0$, ${\mb{E}}_0\parallel {\mb{y}}_0$ and ${\mb{j}}_{2\omega }\parallel {\mb{x}}_0$, similar considerations lead to
\[j_{x2\omega }\approx \frac{ge^4v^2_FE^2_{\omega }E_0}{8\pi {\hbar }^3{\omega }^4}\int^{\pi }_{-\pi }{d{\theta }_{\mb{k}}{\sin}^3{\theta }_{\mb{k}}\cos{\theta }_{\mb{k}}}=0.\] 

%%%%%%%%%%%%%%%%%%%%%%%%%%%%%%%%%%%%%%

\section{Numerical simulations}

We used \cref{GrindEQ__11_,GrindEQ__12_,GrindEQ__13_,GrindEQ__14_,GrindEQ__15_} to simulate the SHG in  graphene illuminated with a strong THz pulse and an optical field beyond the stationary phase approximation. To derive these equations, the amplitudes of optical-frequency coherences  and population differences $\mathrm{\Delta}_{\omega \mb{k}}$ and $\mathrm{\Delta}_{2\omega \mb{k}}$ were assumed to be slow-varying. When the optical field is far off-resonance from an interband transition, Rabi oscillations can have a frequency comparable to the frequency detuning of the optical field. However, the slow-varying assumption is still valid if Rabi oscillations are strongly damped by ultrafast  dephasing processes. So, we use ultrafast population relaxation and dephasing times for hot photoexcited electrons, $\mathrm{\Gamma}^{-1} = 25$ fs  and $\gamma^{-1} = 2.5$ fs, which is consistent in order of magnitude with results from related studies \cite{winzer2010,xing2010,Tan2017} 
 and can be attributed to strong Coulomb interaction between carriers in graphene which results in ultrafast carrier-carrier scattering through interband and intraband Auger recombination and impact ionization \cite{winzer2010,xing2010,Tan2017,tani2012,huang2018}. (Auger recombination could be enhanced due to lattice imperfections in CVD graphene \cite{tani2012}.) Then we can still assume that  the optical-frequency populations and coherences  follow the source terms adiabatically, namely, we can put the $\partial/\partial t$ to be zero in all equations except those for $\mathrm{\Delta}_{0\mb{k}}$. Also, for reasons already discussed above we can assume the optical field to be weak enough to treat it in a perturbative way. 

Another technical difficulty is that $\mathrm{\Omega}_{\mb{k}}$ has a singularity at $|\mb{k}|=0$, which can lead to divergence in numerical simulations. To avoid this problem, we replaced $k$ in the denominator of $\mathrm{\Omega}_{\mb{k}}$ by $k+\epsilon$. We also assumed the chemical potential $\mu_F = 200$ meV and electron temperature at equilibrium $\mathrm{T}_e = 300$ K. The THz field is chosen to be polarized in $y$-direction, and the optical field is polarized in $x$-direction. 

\begin{figure}[htb]
	\begin{center}
		\includegraphics[scale=0.3]{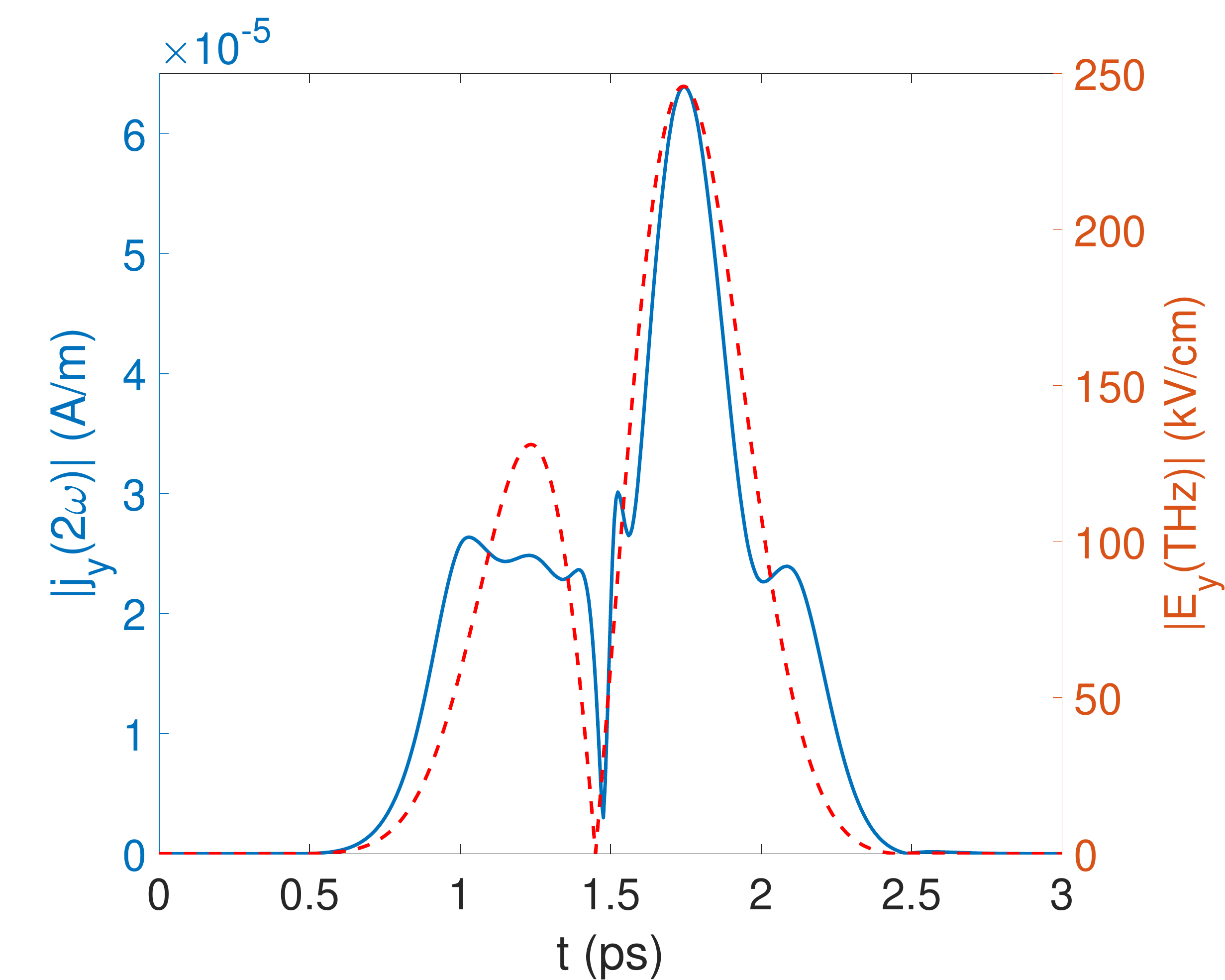}
		\caption{Simulated $y$-component of the SH current (blue solid curve) superimposed onto the profile of the THz field amplitude (red dashed curve). }
		\label{Fig:J_SH_simulation}
	\end{center}
\end{figure}

\begin{figure}[htb]
	\begin{center}
		\includegraphics[scale=0.3]{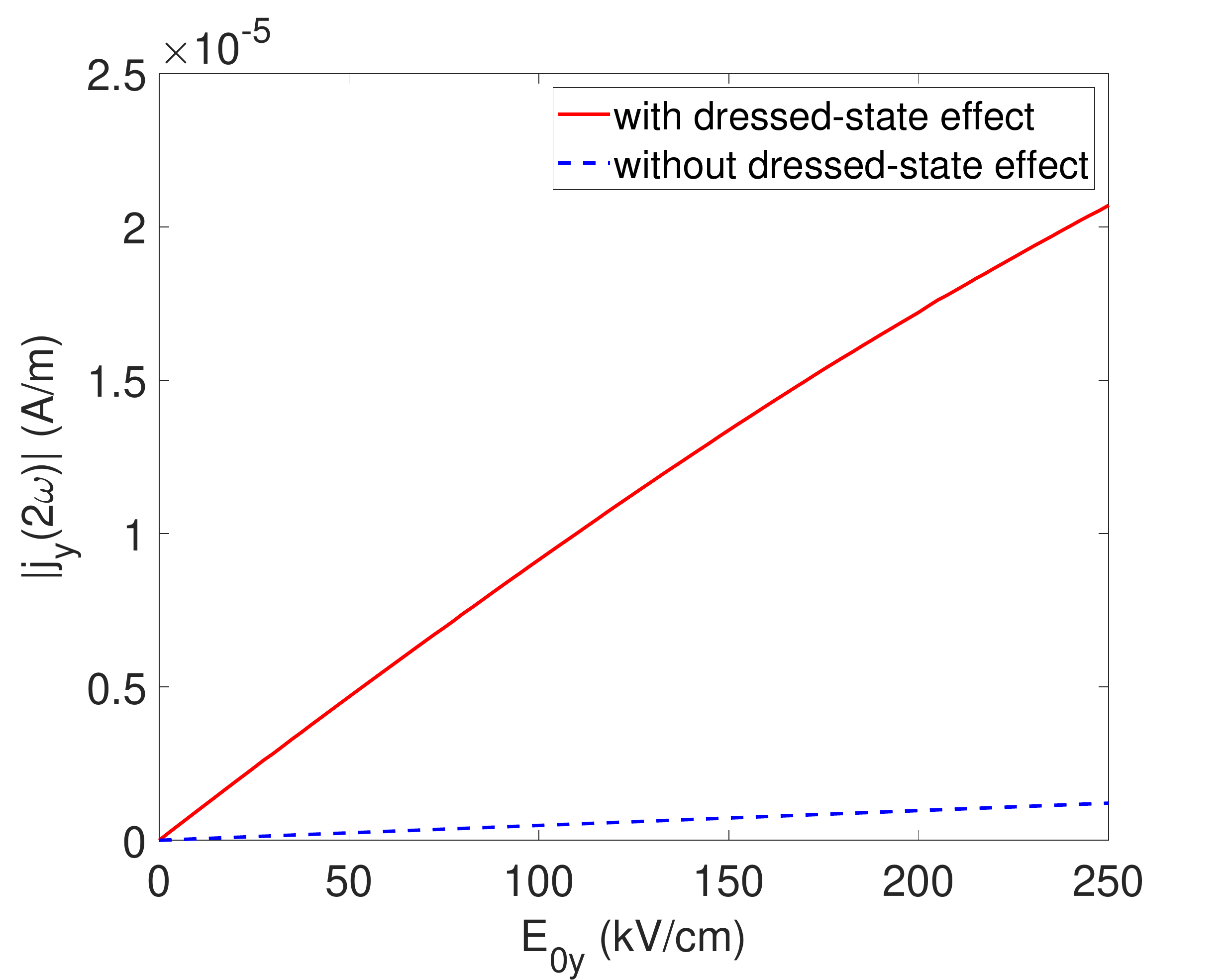}
		\caption{The SH current as a function of the THz field amplitude, calculated for carriers in thermal equilibrium, with (red solid line) and without (blue dashed line) including the dressed-state effect. }
		\label{Fig:J2w_dressed}
	\end{center}
\end{figure}

The simulation shows that the SH current is generated predominantly in $y$-direction, i.e.~along the direction of the THz field. In Fig.~\ref{Fig:J_SH_simulation} we plot the SH current calculated from the simulation, together with the profile of the THz pulse. The SH current generally follows the THz field, except for some small variations originated from the time evolution of the carrier distribution $\mathrm{\Delta}_{0\mb{k}}$. These variations are likely beyond the detector resolution in the experiment. 

Dressing of electron states by the THz field is expected to play an important role in the SHG process, even in the presence of ultrafast scattering. As an illustration, we assume that the carrier distribution $\mathrm{\Delta}_{0\mb{k}}$ is in thermal equilibrium, and calculate the SH current for two cases, with and without $\frac{e \mb{E}_0}{\hbar} \frac{\partial}{\partial \mb{k}}$ terms included. The latter would be similar to four-wave mixing in a two-level medium.  Figure~\ref{Fig:J2w_dressed} shows the dependence of the SH current on the THz field amplituce for the two cases. It indicates that the dressed-state effect can enhance the SH current by one order of magnitude, and therefore the signal intensity by two orders. We can also see that the SH current at the highest THz field of 250 kV/cm is about three times higher in Fig.~5 as compared to Fig.~4 where the thermal equilibrium distribution for $\mathrm{\Delta}_{0\mb{k}}$ was assumed. One could say that roughly 1/3 of the SH current comes from direct parametric interaction between a THz field and an optical field, whereas the nonequilibrium distortion of carrier distribution contributes the remaining 2/3.


%\bibliography{graphene_dressed_SHG_ref}

\end{document}